\documentclass[a4paper,12pt]{article}
\usepackage{epsfig}
\usepackage{color}
\usepackage{amsmath}

\newcommand{\E}{E}
\newcommand{\Eb}{\overline{E}}
\newcommand{\Jb}{\overline{J}}
\newcommand{\Lb}{\overline{L}}
\newcommand{\Db}{\overline{D}}
\newcommand{\chib}{\overline{\chi}}
\newcommand{\Wb}{\overline{W}}

\newcommand{\SKIP}[1]{ }
\allowdisplaybreaks

\begin{document}
\title{Long-range donor-acceptor electron transport mediated by $\alpha$-helices \\
}

\author{
L.S. Brizhik\thanks{e-mail address: brizhik@bitp.kiev.ua},\,\\
Bogolyubov Institute for Theoretical Physics \\
of the National Academy of Sciences of Ukraine, \\ 03143 Kyiv, Ukraine\\
J.  Luo\thanks{e-mail address: j.luo.5@bham.ac.uk}\,\\
School of Mathematics, University of Birmingham, \\ Birmingham B15 2TT, UK\\
B.M.A.G. Piette\thanks{e-mail address: B.M.A.G.Piette@durham.ac.uk},\,
and W.J. Zakrzewski\thanks{e-mail address: w.j.zakrzewski@durham.ac.uk},\,
\\
Department of Mathematical Sciences, University of Durham, \\
Durham DH1 3LE, UK\\
}
\date{}
\maketitle

\begin{abstract}
  We study the long-range electron and energy transfer
mediated by a polaron on an $\alpha$-helix polypeptide chain coupled
to donor and acceptor molecules at opposite ends of the chain. We
show that for specific parameters of the system, an electron initially
located on the donor can tunnel onto the $\alpha$-helix, forming a
polaron, which then travels to the other extremity of the polypeptide
chain where it is captured by the acceptor. We consider three families
of couplings between the donor, acceptor and the chain, and show that
one of them can lead to a 90\% efficiency of the electron transport
from donor to acceptor. We also show that this process remains stable
at physiological temperatures in the presence of thermal
fluctuations in the system. 
   
\end{abstract}
PACS numbers: 05.45.Yv, 05.60.-k, 63.20.kd, 71.38.-k  

\vspace{5mm}

Key words: long-range donor-acceptor electron and energy transfer,
large polaron, alpha-helix, donor-bridge-acceptor complex, soliton,
self-trapping

\section{Introduction}

The mechanisms behind the highly efficient long-range electron transfer
(ET)  in redox reactions accompanying photosynthesis and cellular
respiration have been intensively discussed over several decades
\cite{Jortner, Voet}. This transfer takes place at macroscopic
distances along the so-called electron transport chain in Krebs cycles
in membranes of chloroplasts, mitochondria or cells, and occurs at
physiological temperatures. Conventional mechanisms, such as
tunnelling, Forster and Dexter mechanism etc.~\cite{Forster, Kasha,
Jones}, cannot provide such long-range ET even at zero temperature,
let alone 300 K. Nevertheless, it should be noted that the very
structure of the ET chain can facilitate these processes. An ET chain
consists of a spatially separated sequence of biological molecular
complexes (peptides, enzymes, etc.), along which the sequential
transport of electrons takes place via redox processes, so that
every site in this chain plays the role of an acceptor for the ‘left’
neighbour and donor for the ‘right’ one along the chain (see, e.g.,
\cite{Murray}). The electron transport chain in mitochondria can be
schematically represented as the following sequence: 

$${\rm{NADH+H}}^+ \rightarrow \rm{Complex \ I}  \rightarrow \rm{Q}
\rightarrow \rm{Complex\  II}  \rightarrow \rm{Complex \ III} \rightarrow
$$
$$  \rightarrow \rm{cyt}\ c  \rightarrow \rm{Complex \  IV}
\rightarrow \rm{O}_2 .
$$ 

Here ${\rm{NADH+H}}^+$ is nicotinamide adenine dinucleotide, which
serves as the substrate; Complex I is NADH coenzyme Q reductase; Q is
ubiquinone coenzyme; Complex II is succinate dehydrogenase; Complex
III is cytochrome ${\rm{bc}}_1$; cyt \textit{c} is cytochrome
\textit{c}; Complex IV is cytochrome \textit{c} oxidase; ${\rm{O}}_2$
is molecular oxygen. Another example of the electron transport chain can be found  in \cite{Althoff}.

In each elementary process, at the onset, there is a release
of four electrons at the substrate, which then are carried along the
chain with the reduction of molecular oxygen and hydrogen ions to a
water molecule at the final stage of the process. This transport of
electrons is so exceptionally efficient that only a tiny percentage of
electrons leak out to reduce oxygen.  The
complexes in the ET chain can be conventionally
divided into two groups: heavy and light ones. In particular, in 
ET chains, such elements  as ubiquinone or cytochrome
cyt-\textit{c}, have relatively small molecular weight which leads to
their high mobility. They can move outside the mitochondrial
membrane, carrying electrons from a heavy donor to a heavy acceptor
via a linear, {\it e.g.} the Forster mechanism \cite{Forster,
Jones}. Some other complexes in the electron transport chain, such as
NADH-ubiquinone oxireductase, flavoproteids, cytochrome
\textit{c}-oxidase, cyt –$ aa_3$ and cytochrome cyt –$ bc_1 $
are proteins with large molecular weight of up to several
hundreds of kiloDaltons. Conventional linear mechanisms cannot provide
coherent transport of electrons across these heavy
enzymes, either as a whole or internally between co-factors separated by
macroscopic distances, for instance porphyrins, metal clusters etc.,
that are separated by
macroscopic distance. Nevertheless, their regular crystal-like structure can
facilitate ET, as is discussed below.  For instance, 
inside some large enzymes like NADH ubiquinone oxidoreductase there can be
several long pathways for electron transport \cite{Voet}, where one can identify
the alpha-helical part of the enzyme between the ‘donor’ and the ‘acceptor’.

A significant part of heavy macromolecules is in the alpha-helical conformation,
whose regular structure results in the formation of electron bands in their
energy spectrum. The alpha-helical structure is stabilized by relatively weak
hydrogen bonds resulting in strong electron-lattice interactions, and thus, in
the polaron effect.
An $\alpha$-helical segment of a
protein contains three almost-parallel polypeptide strands bound by
hydrogen bonds along the strands, with weak interactions between these
strands. An isolated strand is described by the Fr\"{o}hlich
Hamiltonian, and this description leads to a system of coupled
nonlinear equations for the electron wavefunction and lattice
variables, and admits soliton solutions. The possibility of
self-trapping of electrons in an isolated one-dimensional molecular
chain, like a polypeptide strand, has been first shown in \cite{DavKisl}
(see also \cite{Davydov85, Scott}) and later it was also demonstrated
in helical systems \cite{DavErSerg, Fed, Brizhik2004}.
The soliton solutions of these models are particular cases of a large polaron.
Such a polaron can be
described as a crossover between an almost-free electron and small polaron
states depending on the strength of the exchange interaction energy,
electron-lattice coupling constant, the number of phonon modes, their type
and the corresponding Debye energies \cite{BE-arb-coup}.
The soliton properties depend on the parameters of
the system. Moreover,  the helical structure of proteins was shown to
lead to the existence of several types of soliton solutions of the model with
different properties and symmetries \cite{Brizhik2004}. In such
soliton states electrons can propagate along macromolecules almost
without any loss of energy. 

The results mentioned above have been obtained for isolated strands or
helices, while in reality, the electron transport occurs in the system
Donor-Bridge-Acceptor, as is the case for the ET chain in the Krebs
cycles. The simple case when the bridge is modelled as a polypeptide
strand had been studied in \cite{Brizhik2014}. It was shown there that
the long-range ET can be provided by the soliton mechanism within a
wide range of parameter values of donor, acceptor and polypeptide
strands. 

In the present paper we study the possibility of a coherent long-range
electron transport in the system Donor-$\alpha$-helix-acceptor. As one
can expect, the formation of the soliton on the $\alpha$-helix depends
on the helix-donnor and helix acceptor coupling, as well as on the parameters
of the system under study (see
{\it e.g.}, \cite{Brizhik2014, Dav-Br-gen, LSB-gen}), and we can find
conditions which lead to the formation of a soliton on the helix.

There are two other aspects of the model developed in the present
paper. The first one is related to the fact that the functioning of
the ET chain is tightly connected with the production of adenosine
triphosphate (ATP): in most organisms the majority of ATP is generated in ET
chains (see, {\it e.g.}, \cite{Nich}). The energy of the hydrolysis of ATP
into ADP
is the basic unit of energy used in biological systems, in particular, in
muscles to produce mechanical work, to establish electrochemical gradients
across membranes, in biosynthetic processes, and in many other physiological
and biochemical processes necessary to maintain life.  The amount of energy
released by ATP hydrolysis is approximately 0.43 eV, which is only 20 times
the thermal energy at physiological temperatures and is not enough for
an electronic excitation. It is sufficient to excite some vibrations, such
as an AMID I vibration, an excitation which requires an energy of 0.21 eV.
AMID I is mainly (up to $80 \% $) the stretching vibration of double C=O bond
of the peptide group which has a relatively large dipole moment 0.3 Db
oriented along the alpha-helix axis. This excitation is registered in optical
spectra of polypeptide molecules, its wavelength being 1650 $ {\rm cm}^ {-1}$,
and, according to \cite{McClare}, the ATP hydrolysis energy is transferred
along protein macromolecules in the form of AMID I vibration.  For more
details see\cite{Nevskaya, DavBiolQM, Davydov85, Scott, Careri1} or
more recently \cite{Ganim, Marques}.

As has been shown by Davydov, the
Amide-I vibration can be self-trapped in macromolecule into a soliton state and
carried along it to the place where it is
utilized for biochemical or mechanical needs \cite{Davydov85,
Scott}. This process, from the mathematical point of view, is
described formally by the same system of equations as the
ET. Therefore, the results obtained here are equally valid for such
energy transfer processes. 

The second aspect of the model is related to the potential importance
of our results for micro- and nano-electronics where conjugated
donor-acceptor copolymer semiconductors with intra-molecular charge
transfer on large distances are widely used. A large number of such
systems have been recently synthesized. They include donor-acceptor
pairs mediated by salt bridges \cite{Roberts}, thienopyrazine-based
copolymers \cite{Champion} and some others \cite{Li2, Nguyen,
Tian}. Donor-bridge-acceptor systems with efficient ET play an important
role in electronic applications \cite{Li1, vanM, Zhu, Yu}: they can be
used in photovoltaic cells \cite{Campos, Svenss, Adm, Tian},
light-emitting diodes \cite{Kulkarni, Ego, Thomp, Wu} and field-effect
transistors \cite{Babel, Yamamoto, Yasuda, ChenLee}, in particular,
thin-film organic field effect transistors \cite{Zhang}. Proteins and
synthetic macromolecules have a great technological potential; one 
example is the
improvement of efficiency and UV-photostability of planar perovskite
solar cells using amino-functionalized conjugated polymers as ET
materials \cite{ Li2,Tian}. 

Recent novel applications in bioelectronics such as organic photovoltaics,
fuel cell technology and other, are based on metal–organic frameworks or
structures, that are complexes of electroconducting compounds/substrates and
polypeptides (see, e.g., \cite{Chen, Sepunaru, Mileo} and references therein).
It has been shown that both the peptide composition and structure can affect
the efficiency of electron transport across peptides \cite{Sepunaru}. Moreover,
long-range conductivity and enhanced solid-state electron transport in
proteins and peptide bioelectronic materials has been proven experimentally
\cite{Ing, Amdursky}. The effectiveness of electron transport processes in
living systems is already used in novel electronic devices, e.g.,  in
Shewanella Oneidensis MR-1 Cells, based on multiheme cytochrome mediated
redox conduction \cite{Xu}, or in synthesized supramolecular charge transfer
nanostructures based on peptides \cite{ Nalluri}, synthetic biological protein
nanowires with high conductivity \cite{Tan}, self-assembled peptide nanotubes
used as electronic aterials \cite{Akdim} and many others.  We quote \cite{Ing}:
“The ability of such natural and synthetic protein and
peptide materials to conduct electricity over micrometer to centimeter length
scales, however, is not readily understood from a conventional view of their
amino acid building blocks. Distinct in structure and properties from
solid-state inorganic and synthetic organic metals and semiconductors,
supramolecular conductive proteins and peptides require careful theoretical
treatment…”. This is one of the factors which have motivated our interest in
the problems discussed in the present paper and we hope that our study will
shed some light on this problem.

In the first section of the paper we derive a model of the
$\alpha$-helix coupled to a donor molecule and an acceptor molecule. This model
is a combination of the models derived in \cite{Brizhik2004} and
\cite{Brizhik2014}. We then perform a parameter scaling to make all
the parameters dimensionless and derive the equations in such
units. After selecting the parameters that best describe the
$\alpha$-helical protein, we compute the profile of a static
self-trapped electron state (soliton-like or, in other words, large
polaron state, which for simplicity we call from now on a `polaron')
by solving the model equations numerically. We then study various
configurations where the electron density has been set to 1 on the
donor and 0 elsewhere and let the system evolve. We do this for three
different types of couplings between the donor and acceptor to the
$\alpha$-helix and we determine  numerically the donor and acceptor
coupling parameters that lead to the best transfers of the
electron. We end the paper by describing the solutions we have found
and draw some conclusions.

\section{Model of the System `Donor -- $\alpha$-Helix -- Acceptor' }
\begin{figure}[!ht]
    \centering 
    \includegraphics[width=14cm]{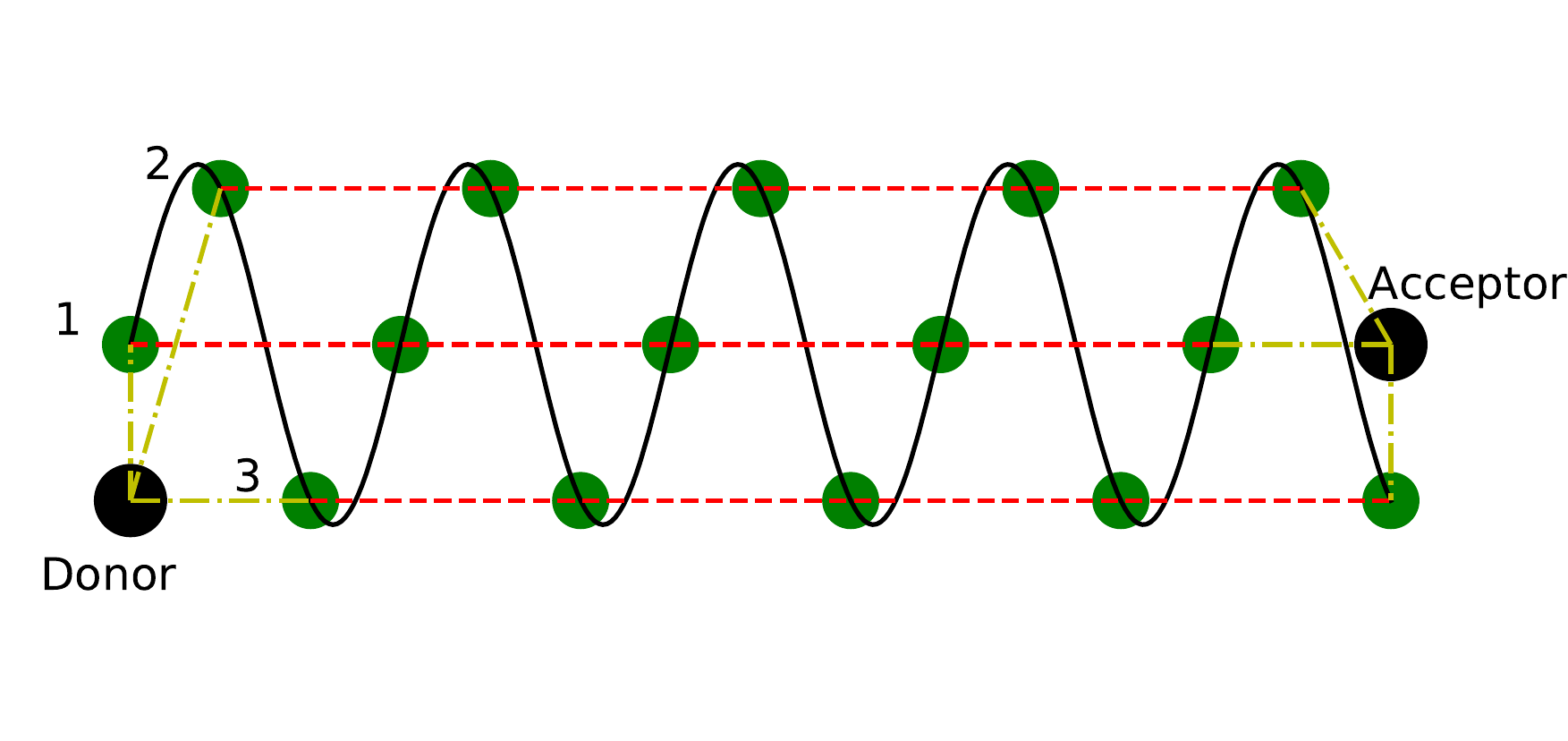}
    \vspace{-15mm}
    \caption{ The model of $\alpha$-helix with a donor and an acceptor.
   The continuous line represents the helix backbone formed by chemical bonds, the dash lines represent the hydrogen bonds that are links along the strands and the dash-dot lines the links
   between the donor/acceptor and the different strands.
   The numbers 1, 2 and 3 label the 3 strands.
}
\label{fig:helix}
\end{figure}

We consider a polypeptide chain in an $\alpha$-helical configuration made out
of $N$ peptide  groups (PGs), with a donor molecule attached to one end and
an acceptor
molecule attached to the other end. The peptide chain forms a helical
structure in which each molecule is coupled by chemical bonds to its
neighbours along the chain as well as to the PG 3 sites away from it by
hydrogen bonds. With this 3-step coupling, the
$\alpha$-helix can also be seen as 3 parallel chains~\cite{Davydov74} which we
refer to as strands in what follows. This model is depicted in Fig. 
\ref{fig:helix}.

We label the PGs with the index $n$
along the polypeptide chain, and use $n=0$ for the donor and $n=N+1$ for the
acceptor. This means that PGs with an index difference which is a
multiple of 3 belong to the same strand of the $\alpha$-helix.

The donor and the acceptor can, {\it a-priori}, be coupled respectively to the
first 3 and
the last 3 peptides, {\it i.e.}, with the nodes $n=1,2,3$ and $N=N-2, N-1, N$.
In our study, we will consider 3 different types of couplings but for now,
we assume that all the coupling parameters are different.

The Hamiltonian of the system is given by 
\begin{eqnarray}
  {\cal{H}}\, = {\cal{H}}_e\,+\,{\cal{H}}_p\,+\,{\cal{H}}_{int} ,
  \label{eq:H}
\end{eqnarray}
where ${\cal{H}}_e$, ${\cal{H}}_p$ and ${\cal{H}}_{int}$ are respectively the
phonon, electron and interaction Hamiltonians given by 
\begin{align}
  {\cal{H}}_e\, &= \,\Eb_d\,|\Psi_{0}|^2\,+ \,\Eb_a\,|\Psi_{N+1}|^2\,
           +\,\Eb_0\,\sum_{n=1}^N\,|\Psi_{n}|^2\,
       -\,\Jb\,\sum_{n=1}^{N-3}\,\Big(\Psi_{n}\Psi_{n+3}^*\,+\,\Psi_{n+3}\Psi_{n}^*\Big)\nonumber\\  
     &\quad +\,\Lb\,\sum_{n=1}^{N-1}\,\Big(\Psi_{n}\Psi_{n+1}^*\,+\,\Psi_{n+1}\Psi_{n}^*\Big)
\,-\,\sum_{\ell=1}^3 \Db_{d,\ell}(\Psi_{0}\Psi_{\ell}^*+\Psi_{\ell}\Psi_{0}^*) \nonumber\\
&\quad -\,\sum_{\ell=1}^3 \Db_{a,\ell}(\Psi_{N+1}\Psi_{N-3+\ell}^*+\Psi_{N-3+\ell}\Psi_{N+1}^*) , \\
{\cal{H}}_p\, &=\,\frac{1}{2} \Bigl[{P_{d}^2\over M_d}\,+\,{P_{a}^2\over M_a}\,
                  \Bigl]\,+\,\frac{1}{2}\,\sum_{\ell=1}^3 \Big[\Wb_{d,\ell} (U_{0} - U_{\ell})^2+\Wb_{a,\ell} (U_{N+1} - U_{N-3+\ell})^2\,\Big]
                  \nonumber\\
                  &\quad \,+\,  \frac{1}{2}  \sum_{n=1}^N {P_{n}^2\over M}\,
                    +\,\frac{1}{2}\,\sum_{n=1}^{N-3} \Wb (U_{n+3} - U_{n})^2 , \\
  {\cal{H}}_{int}\, &= |\Psi_0|^2\,\sum_{\ell=1}^3 \,\chib_{d,\ell}(U_\ell-U_0)
       +\,|\Psi_{N+1}|^2\,\sum_{\ell=1}^3 \,\chib_{a,\ell}(U_{N+1}-U_{N-3+\ell})\,
                        \nonumber\\
                    &\quad +\sum_{\ell=1}^3 \,|\Psi_\ell|^2\left[\,\chib_{d,\ell}\,(U_\ell-U_0)
+\,\chib\,(U_{\ell+3}-U_\ell)\right]
                  \nonumber\\
           &\quad +\,\sum_{\ell=1}^3 \,|\Psi_{N-3+\ell}|^2\,\left[\chib_{a,\ell}\,(U_{N+1}-U_{N-3+\ell})\,
+\,\chib\,(U_{N-3+\ell}-U_{N-6+\ell})\right]
\nonumber\\
&\quad + \,\chib\,\sum_{n=4}^{N-3} |\Psi_{n}|^2(U_{n+3} - U_{n-3}).
\label{Hint_al}
\end{align}



In these expressions, $\Eb_0$ describes the on-site electron energy, $\Jb$ the
resonance integral along the strands, $\Lb$ the resonance
integral along the helix, $M$ the mass of the unit cell, $\chib$ the
electron-lattice coupling and $\Wb$ the elasticity of the bond along the
strands. The constants with subscript $d$ and $a$ refer to parameters of
the donor and the acceptor respectively.

The functions $\Psi_{n}$ describe the electron wave function (and so $\vert\Psi_n\vert^2$ describe the electron probability of being at the site $n$)
and $U_n$ describe the displacement of molecule $n$ along the strands. 
$P_n$ are the canonically conjugated momenta of $U_n$.
Of course, the electron wave function satisfies the 
normalization condition
\begin{eqnarray}
\sum_{n=0}^{N+1} |\Psi_n|^2 = 1,
\label{eq_norm}
\end{eqnarray}
where, following our convention, $\Psi_0=\Psi_d$ and $\Psi_{N+1}=\Psi_a$.

Our model is meant to describe the case in which the principal chain can be
sufficiently well approximated by one phonon band corresponding to an acoustical
phonon mode which describes the longitudinal displacements
of the unit cells from their positions of equilibrium along the helix's strands.
The electron-lattice interaction Hamiltonian induces a dependence of the 
electron Hamiltonian on the
lattice distortions. We also assume here that the dependence of the on-site
electron energy on the lattice distortion is much stronger than that of
the inter-site electron interaction energy.

The model we present here is a combination of the polaron model of the
$\alpha$-helix which was described in detail in \cite{Brizhik2004} and of
the donor-acceptor model described in \cite{Brizhik2014}. The first model
describes polarons on an $\alpha$-helix, instead of using the traditional
single chain, proposed by Davydov~\cite{Davydov85,Scott}, which corresponds
to what we call a strand in this paper. In fact, it was shown in
\cite{Brizhik2004} that the polaron  is spread over the 3 strands hence the
relevance of using a more realistic helical model.
The second paper describes a
model of the transfer of an electron from a donor molecule to
an acceptor one via the coherent propagation of a polaron along a simple chain
(a single strand in the present model). The model we describe here is a
combination of these two models in which the donor and the acceptor
are coupled to a proper $\alpha$-helix instead of to a single strand.

\section{Parameter scaling}
To facilitate the analysis of the model solutions, it is convenient
to scale the parameters so that they become dimensionless. Thus, following
\cite{Brizhik2004}, we perform the following scalings:
\begin{eqnarray}
\begin{array}{llll}  
   d = 10^{-11} \textnormal{ m},\qquad 
  &u_n = \frac{U_n}{d},\qquad 
  &\tau = t\nu, \\
   \E_0= \frac{\Eb_0}{\hbar\nu},\qquad 
  &\E_d = \frac{\Eb_d}{\hbar\nu},\qquad 
  &\E_a = \frac{\Eb_a}{\hbar\nu},\\
   J=\frac{\Jb}{\hbar\nu},\qquad 
  &D_a = \frac{\Db_a}{\hbar\nu},\qquad 
    &D_d = \frac{\Db_d}{\hbar\nu},\\
   W = \frac{\Wb}{\nu^2\,M}, \qquad 
  &W_{d,\ell} = \frac{\Wb_{d,\ell}}{\nu^2\,M}, \qquad 
  &W_{a,\ell} = \frac{\Wb_{a,\ell}}{\nu^2\,M}, \\
   \chi = \frac{d\,\chib}{\hbar\, \nu},\qquad 
  &\chi_{d,\ell} = \frac{d\,\chib_{d,\ell}}{\hbar\, \nu}, \qquad 
  &\chi_{a,\ell} = \frac{d\,\chib_{a,\ell}}{\hbar\, \nu}, \\
L=\frac{\Lb}{\hbar\nu},\qquad    
   &K_d = \frac{M}{M_d},\qquad 
  &K_a = \frac{M}{M_a}.
\end{array}    
\end{eqnarray}
As a result, the Hamiltonian takes the form 
${\cal{H}}_p= \, M\, \nu^2\, d^2\,H_p$,
${\cal{H}}_e\,=\,\hbar\, \nu\,\,H_e$ and
${\cal{H}}_{int}= \,\hbar\, \nu\, \,H_{int}$ where the dimensionless terms are 
\begin{align}
  H_e\, &=\,\E_d\,|\Psi_{0}|^2\,+ \,\E_a\,|\Psi_{N+1}|^2\,
           +\,\E_0\,\sum_{n=1}^N\,|\Psi_{n}|^2\,
       -\,J\,\sum_{n=1}^{N-3}\,\Big(\Psi_{n}\Psi_{n+3}^*\,+\,\Psi_{n+3}\Psi_{n}^*\Big)\nonumber\\
  &\quad +\,L\,\sum_{n=1}^{N-1}\,\Big(\Psi_{n}\Psi_{n+1}^*\,+\,\Psi_{n+1}\Psi_{n}^*\Big)
     \,
     -\,\sum_{\ell=1}^3 D_{d,\ell}(\Psi_{0}\Psi_{\ell}^*+\Psi_{\ell}\Psi_{0}^*)
     \nonumber\\
        &\quad \,-\,\sum_{\ell=1}^3 D_{a,\ell}(\Psi_{N+1}\Psi_{N-3+\ell}^*+\Psi_{N-3+\ell}\Psi_{N+1}^*) , \label{eq:He} \\
  H_p\, &=\,\frac{1}{2} \Bigl[\frac{1}{K_d}\left(\frac{du_{0}}{dt}\right)^2\,
            +\,\frac{1}{K_a}\left(\frac{du_{N+1}}{dt}\right)^2\,
            \Bigl]\, +
            \nonumber\\
                      &\quad \, +\,\frac{1}{2}\,\sum_{\ell=1}^3 \Big[W_{d,\ell} (u_{0} - u_{\ell})^2
            +W_{a,\ell} (u_{N+1} - u_{N-3+\ell})^2\,\Big]
                  \nonumber\\
         &\quad \,+\, \frac{1}{2}\sum_{n=1}^N \left(\frac{du_{n}}{dt}\right)^2\,
                    +\,\frac{1}{2}\,\sum_{n=1}^{N-3} W (u_{n+3} - u_{n})^2,  \label{eq:Hp} \\
  H_{int}\, &= |\Psi_0|^2\,\sum_{\ell=1}^3 \,\chi_{d,\ell}(U_\ell-U_0)
       +\,|\Psi_{N+1}|^2\,\sum_{\ell=1}^3 \,\chi_{a,\ell}(U_{N+1}-U_{N-3+\ell})\,
                        \nonumber\\
          &\quad +\sum_{\ell=1}^3 \,|\Psi_\ell|^2\left[\,\chi_{d,\ell}\,(U_\ell-U_0)
+\,\chi\,(U_{\ell+3}-U_\ell)\right]
                  \nonumber\\
           &\quad +\,\sum_{\ell=1}^3 \,|\Psi_{N-3+\ell}|^2\,\left[\chi_{a,\ell}\,(U_{N+1}-U_{N-3+\ell})\,
+\,\chi\,(U_{N-3+\ell}-U_{N-6+\ell})\right]
\nonumber\\
&\quad +\,\chi\,\sum_{n=4}^{N-3} |\Psi_{n}|^2(U_{n+3} - U_{n-3}).
           \label{eq:Hint}
\end{align}

We must thus have $M\,\nu^2\, d^2 = \hbar\, \nu$ and so
$\nu\,=\,\hbar/(M\,d^2)$. With $M=1.9112\times 10^{-25}$ kg \cite{Brizhik2014}
and, as $\hbar=1.054\times 10^{-34}$ Js, we have $\nu= 5.51 \times 10^{12}$
s\textsuperscript{$-1$}.

Before deriving the dimensionless equations it is also convenient to multiply
the wave function by a time-dependent phase and so we define 
\begin{eqnarray}
\psi(t) =  \Psi(t)\exp\left(-\frac{it}{\hbar}(\Eb_0+2\Lb-2\Jb)\right).
\end{eqnarray}

Following \cite{Brizhik2014} we also add to the acceptor equation a term of
the form $i \sum_{\ell=1}^3 A_{a,\ell} |\psi_{N-3+\ell}|^2\psi_{N+1}$, which
describes the transfer of the electron from the alpha-helix to the acceptor
and has a clear physical meaning:  the higher the probability of the electron
localization at the terminal end of the helix, the higher the probability
of its transfer to the acceptor. It is easy to check that this extra term
does not violate conservation of the total electron probability.

From the above Hamiltonian (\ref{eq:H}),(\ref{eq:He})-(\ref{eq:Hint})
one can easily derive the following equations for $U_n$ and $\Psi_n$:
\begin{eqnarray}
  i\frac{d\Psi_{0}}{d\tau} &=& (\E_d-\E_0-2L+2J)\Psi_{0}-\sum_{\ell=1}^3 D_{d,\ell}\Psi_{\ell}
  +\Psi_{0}\sum_{\ell=1}^3\chi_{d,\ell}(u_\ell-u_0),\nonumber\\
  i\frac{d\Psi_{\ell}}{d\tau} &=&  (2J-2L) \Psi_{\ell} - J \Psi_{\ell+3}+L(\Psi_{\ell+1}+\Psi_{\ell-1}(1-\delta_{\ell,1}))
                            -D_{d,\ell}\Psi_{0}\nonumber\\
                           && +\chi_{d,\ell}\Psi_{\ell}(u_\ell-u_0)
                              +\chi \Psi_{\ell}(u_{\ell+3}-u_{\ell}),
                              \qquad l=1,2,3,\nonumber\\
  i\frac{d\Psi_{n}}{d\tau} &=& (2J-2L)  \Psi_{n} -J (\Psi_{n+3}+\Psi_{n-3})
                            +L(\Psi_{n+1}+\Psi_{n-1})
                            +\chi \Psi_{n}(u_{n+3}-u_{n-3}),\nonumber\\
                           && \qquad n=4 \dots N-3, \nonumber\\
  i\frac{d\Psi_{N-3+\ell}}{d\tau} &=&(2J-2L)  \Psi_{N-3+\ell} -J \Psi_{N-6+\ell}
                            +L(\Psi_{N-4+\ell}+\Psi_{N-2+\ell}(1-\delta_{\ell,3}))
                            -D_{a,\ell}\Psi_{N+1}\nonumber\\
                            &&+\chi_{a,\ell}\Psi_{N-3+\ell}(u_{N+1}-u_{N-3+\ell})
                               +\chi \Psi_{N-3+\ell}(u_{N-3+\ell}-u_{N-6+\ell})
                               \nonumber\\
                           &&-i A_{a,\ell} |\Psi_{N+1}|^2\Psi_{N-3+\ell},
                        \qquad l=1,2,3,\nonumber\\
  i\frac{d\Psi_{N+1}}{d\tau} &=& (\E_a-\E_0+2J-2L) \Psi_{N+1}-\sum_{\ell=1}^3 D_{a,\ell}\Psi_{N-3+\ell}
                                 +\Psi_{N+1}\sum_{\ell=1}^3\chi_{a,\ell}(u_{N+1}-u_{N-3+\ell})\nonumber\\
                           &&+i \sum_{\ell=1}^3 A_{a,\ell} |\Psi_{N-3+\ell}|^2\Psi_{N+1},
 \nonumber\\
  \frac{d^2 u_{0}}{d\tau^2}&=& K_d\Big(\sum_{\ell=1}^3 W_{d,\ell}(u_\ell-u_0) 
                               +\sum_{\ell=1}^3 \chi_{d,\ell}(|\Psi_0|^2+|\Psi_\ell|^2)\Big),
                            \nonumber\\ 
  \frac{d^2 u_{\ell}}{d\tau^2}&=& W(u_{\ell+3}-u_\ell) + W_{d,\ell}(u_0-u_\ell)
                            -\,\chi_{d,\ell}(|\Psi_0|^2+|\Psi_\ell|^2)
                                  +\,\chi(|\Psi_\ell|^2+|\Psi_{\ell+3}|^2),
                   \nonumber\\ &&  \qquad \qquad \qquad \ell=1,2,3,\nonumber\\ 
  \frac{d^2 u_{n}}{d\tau^2}&=& W(u_{n+3}+u_{n-3}-2u_n)
                            +\,\chi\,(|\Psi_{n+3}|^2-|\Psi_{n-3}|^2),
                               \qquad n=4 \dots N-3,\nonumber\\ 
  \frac{d^2 u_{N-3+\ell}}{d\tau^2}&=& W(u_{N-6+\ell}-u_{N-3+\ell}) 
                               + W_{a,\ell}(u_{N+1}-u_{N-3+\ell})\nonumber\\ 
                          &&+\,\chi_{a,\ell}(|\Psi_{N+1}|^2+|\Psi_{N-3+\ell}|^2)
                             -\,\chi(|\Psi_{N-3+\ell}|^2+|\Psi_{N-6+\ell}|^2),
                             \qquad l=1,2,3,\nonumber\\ 
\frac{d^2 u_{N+1}}{d\tau^2}&=& K_a\Big(\sum_{\ell=1}^3 W_{a,\ell}(u_{N-3+\ell}-u_{N+1}) 
                            -\sum_{\ell=1}^3 \,\chi_{a,\ell}(|\Psi_{N+1}|^2+|\Psi_{N-3+\ell}|^2)\Big),
\label{eq:scaled}
\end{eqnarray}
where $\delta_{i,j}$ is the Kronecker delta function. We now need to select the parameter values that best describe the
$\alpha$-helix.


\subsection{Parameter values}

For the numerical modelling we need to use some numerical values of
the parameters. We recall that, in particular, the  parameter values
for the polypeptide macromolecules are:  $J_\textnormal{Amide-I}= 1.55
\times 10^{-22}$ Joules $\approx 10^{-3}$ eV;  $J_e\approx 0.1-0.01$ eV
$\approx 10^{-21}-10^{-20}$ Joules; $\chi=(35 - 62)$ pN; $w=39 - 58$
N/m, $V_{ac}=(3.6-4.5) \times 10^3$ m/s \cite{Scott}.  The molecular
weights of large macromolecules which participate in the electron
transport chain in redox processes are: NADH-ubiquinone oxidoreductase
- 980 kDa; cytochrome  $bc_1$ complex - 480 kDa; cytochrome $c-aa_3$
oxidase - 420 kDa. The mass of  Cyt-c    is 12 kDa, in which the hem-A
group has a molecular weight 852 Da, and  hem-B group has 616 Da,
which are 3-5 times larger than the molecular weight, 100-200 Da,  of
amino-acids that form macromolecules. Studies of the 
mitochondrial ET chain shows that the electrochemical potential for the
transfer of an electron is $E_{e-c} = +1.135$ V   \cite{Engel2007, Collini2010}.

For completeness of the study we also summarize the data on the parameter
values of other relevant compounds in accordance with the discussion in the
introduction. The molecular weights of many conjugated polymer semiconductors
vary in the interval (10 - 176) kDa, and the hole mobility is
$4 \times  10^{-4} - 1.6 \times 10^{-3}\, {\rm cm}^2$ /(V s).
The ionization potential and electron affinity potential for some donor-acceptor
copolymer semiconductor molecules are: (2.5-4.5) eV 
and (1.5-3.1) eV, respectively \cite{Renger2001}. 
The electrochemical band gap is $E_g^{(el)} = E_{IP} - E_{EA}$ is 
 1.5 eV for BTTP, 1.84 eV for BTTP-P, and 2.24
eV for BTTP-F, which are 0.4-0.6 eV larger from the optically
determined ones $E_g^{(opt)}= 1.1-1.6 $ eV. This difference can be
explained by the exciton binding energy of conjugated polymers
which is thought to be in the range of $E_{ex} \approx 0.4-1.0 $ eV
\cite{Wang2007}. Thieno pyrazine-based donor-acceptor copolymers, such as BTTP, 
BTTP-T, BTTP-F, BTTP-P, have moderate to high molecular weights,
broad optical absorption bands that extend into the near-infrared region with absorption maxima at 667-810 nm,
and small optical band gaps (1.1 - 1.6 eV). They show ambipolar redox properties with 
low ionization potentials
(HOMO levels) of (4.6--5.04) eV. The field-effect mobility of holes varies 
from $4.2 \times 10^{-4}\, {\rm cm}^2$/(V s) in BTTP-T
to $1.6 \times 10^{-3} {\rm cm}^2$/(V s) in BTTP-F (see \cite{Champion}). 
The reduction potentials of BTTP, BTTP-P, and BTTP-F are
-1.4, -1.73, and -1.9 V (vs SCE), respectively. The 
oxidation potentials of the copolymers are in the range 0.29-0.71 V (vs SCE).
The onset oxidation potential and onset reduction potential
of the parent copolymer BTTP are 0.2 and -1.3 V, respectively,
which give an estimate for the ionization potential (IP, HOMO
level) of 4.6 eV ($E_{IP} = E_{ox}^{onset} + 4.4)$  and an electron affinity
(EA, LUMO level) of 3.1 eV ($E_{EA} = E_{red}^{onset} + 4.4$). The 4.6
eV $E_{IP}$ value of BTTP is 0.3 eV lower than that of poly(3-
hexylthiophene) (4.9 eV), whereas its $E_{EA}$ value (3.1 eV) is 0.6
eV higher than that reported for the poly(2,3-dioctylthieno[3,4-b]pyrazine) 
homo-polymer ($\approx 2.5$ eV). An $E_{IP}$ value of 4.64 eV and $E{EA}$ value of 2.8 eV were found in the case of BTTP-P  \cite{Champion}.

In what follows, we set the
on-site electron energy level as the zero of energy, hence we take
$\Eb_0= 0\,$.
We are also using a set of model parameters close to those
encountered in polypeptide macromolecules or to the bridge-mediated
donor-acceptor systems summarized above {\it i.e.}
\begin{equation}
    \Jb=8.42\times 10^{-23}\,\rm{J}, \quad
   \Lb=1.34\times 10^{-22}\,\rm{J}, \quad 
   \Wb  = 10.59\,\rm{kg}/\rm{s}^{2}, \quad
    \chib = 1.85\times 10^{-11}\,\rm{J}/m
  \label{eq:paramsdim}
\end{equation} 
corresponding to the following adimensional values of the parameters in
our equations
\begin{equation}
    J=0.145, \,
  L=0.231, \,
  W = 1.825, \,
  \chi = 0.318.
  \label{eq:params}
\end{equation} 

The order of magnitude of these parameters values is close to the parameter
values
for the electron transport in polypeptides and for other systems described
above. Our aim, for these systems, is to establish a proof of concept
of the soliton mediated long-range ET rather than a performing a detailed
study of their actual fine properties.

\begin{figure}[!ht]
    \centering 
 \includegraphics[width=60mm]{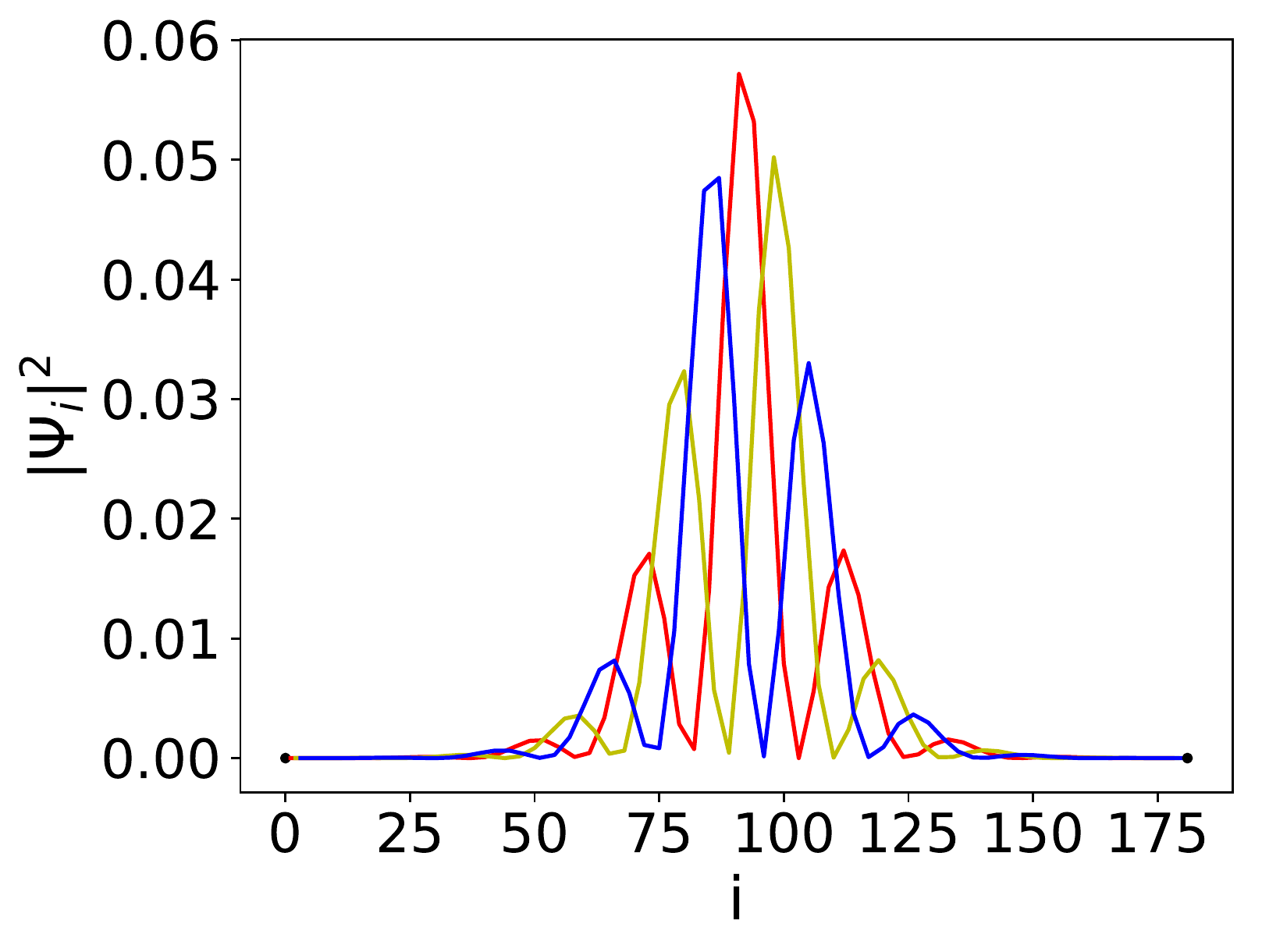}
 \includegraphics[width=60mm]{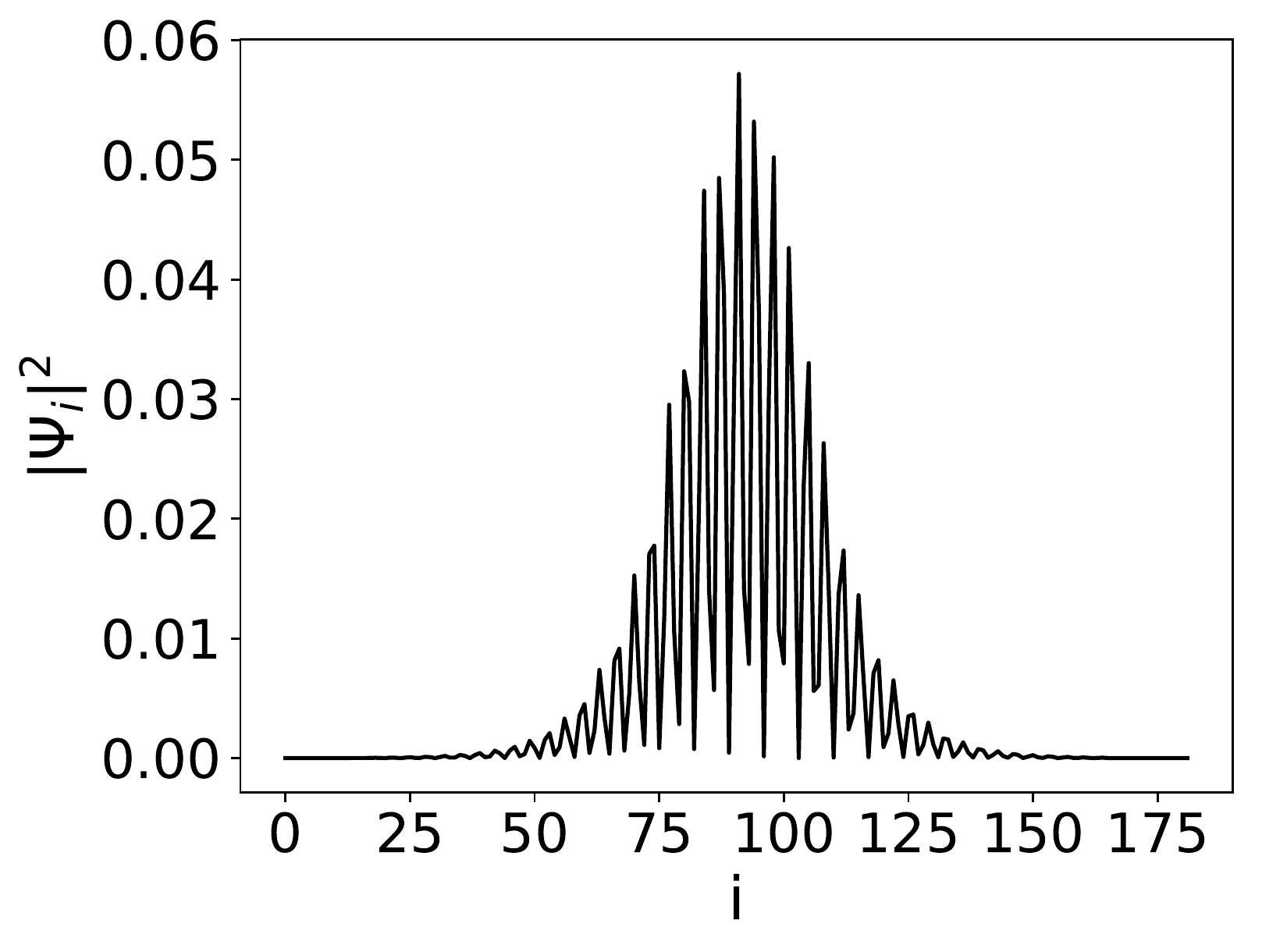}

 a\hspace{60mm} b
 
 \caption{Polaron with $\E_0=\E_d=\E_a=0$,
   $J=0.145$, $L=D_{d,3}=D_{a,1}=0.231$, $D_{d,1}=D_{d,2}=D_{a,2}=D_{a,3}=0$,  
   $W=W_{d,1}=W_{d,2}=W_{d,3}=W_{a,3}=1.825$, $W_{a,1}=W_{a,2}=0$,
   $\chi_{d,\ell}=0.318$, $\chi_{d,\ell} = \chi_{a,\ell} = 0$, $A_{a,ell}=0$,
   $K_d = K_a=1$. a) The electron probability densities are plotted versus the
   index on the polypeptide chain. The 3 strands profiles are shown as
   separate curves. b) The electron density along the $\alpha$-helix
   backbone. 
}
\label{fig:pol_relax}
\end{figure}

Before studying the transfer of an electron from the donor to the
acceptor we have computed the profile of the static polaron on the helix
for the parameters given in (\ref{eq:params}).
This profile is shown on Fig. \ref{fig:pol_relax}. To obtain this profile,
we have relaxed the equations (\ref{eq:scaled}), using donor-acceptor
parameter values so that they do not interact with the chain.

One sees clearly from Fig. \ref{fig:pol_relax}, where the index $i$ runs
along the polypeptide helix and where each curve corresponds to a different
strand, that the static polaron is a broad
localised lump which winds around the polypeptide chain rather than a
single soliton located on a single strand or three identical solitons located
on each of the strands.

\section{Classes of Couplings}
Having so far defined a model with a general set of couplings between the
$\alpha$-helix and the donor and acceptor, we will now restrict ourselves
to 3 families of couplings. 

In the first set, the donor and the acceptor are coupled to all 3
strands of the helix using identical coupling parameters. So we have
\begin{eqnarray}
  D_{d,1}=D_{d,2}=D_{d,3}\neq 0,&&\qquad D_{a,1}=D_{a,2}=D_{a,3}\neq 0,\nonumber\\
  W_{d,1}=W_{d,2}=W_{d,3}\neq 0,&&\qquad W_{a,1}=W_{a,2}=W_{a,3}\neq 0.
\end{eqnarray}
We call such a configuration the `full homogeneous' coupling.

The second configuration describes the case in which the donor and the
acceptor are coupled to only one strand, so that 
\begin{eqnarray}
  D_{d,1}\neq 0,\quad D_{d,2}=D_{d,3}=0,&&\quad D_{a,1}\neq 0,\quad D_{a,2}=D_{a,3}=0,\nonumber\\
  W_{d,1}\neq 0,\quad W_{d,2}=W_{d,3}=0,&&\quad W_{a,1}\neq 0,\quad W_{a,2}=W_{a,3}=0,\nonumber\\
  A_{a,1}\neq 0,&&\qquad A_{a,2}=A_{a,3}=0.
\end{eqnarray}
We call this the `single strand' coupling. Notice that the donor is coupled
to the first peptide of the helix, {\it i.e.}, to the first peptide group of
the first strand, but the acceptor is coupled to the second
but last peptide of the helix, {\it i.e.}, to the last peptide group on the
same strand.

For the third configuration we consider the case when the donor and the
acceptor are coupled only to the
first and last peptides on the alpha-helix so 
\begin{equation}
D_{d,1}\neq 0,\quad D_{a,3}\neq 0,,\quad W_{d,1}\neq 0,\quad W_{a,3}\neq 0, \quad
 A_{a,3}\neq 0, 
\end{equation}
while the other parameters are equal to zero.
We call this case the `end to end' coupling.

To find the best parameter values for the transfer of the electron from the
donor to the acceptor, we have integrated the system of equations
\eqref{eq:scaled} numerically on a lattice of 180 PGs. As the initial
condition we have set the electron probability density to $1$ on the
donor and to $0$ everywhere else. We then
integrated the equations (\ref{eq:scaled}) numerically up to $\tau=500$.
This time was so chosen because it is roughly 3 times longer than it takes for the
polaron to reach the end of the 180-peptides chain.
The value of $|\Psi_{N+1}|^2$
varies with time, but tends to increase modulo some oscillations. To evaluate
$\max |\Psi_{N+1}|^2$ we have tracked its value during the evolution and
recorded the largest value obtained before $\tau \le 500$.

We have first determined the best donor parameters so that the
electron is fully transferred onto the $\alpha$-helix.
We then scanned a very large range of parameter values for the acceptor
to determine the one for which the maximum value of the electron probability
density on the acceptor, $\max |\Psi_{N+1}|^2$, reaches the largest value. 

We will now describe the results  we have obtained for each type of coupling. 

\subsection{Full Homogeneous Coupling}

 The best parameter values we have found to generate a transfer of electron
from the donor to the acceptor are (assuming all the values of $A_{a,\ell}$,
$D_{d,\ell}$, $D_{a,\ell}$ $W_{d,\ell}$, $W_{a,\ell}$, $\chi_{d,\ell}$ and
$\chi_{a,\ell}$  are the same for $\ell=1,2,3$):
\begin{eqnarray}
  &&\E_d=0.25, \quad D_{d,\ell}=0.38\,J, \quad W_{d,\ell}=0.32\,W, \quad \chi_{d,\ell}=0.62\,\chi. \label{eq:best_hc}\\
  &&A_{a,\ell}=0.62, \quad \E_a=0.194,\quad D_{a,\ell}=0.175\,J, \quad
     W_{a,\ell}=0.14\,W,\quad  \chi_{a,\ell}=0.27\,\chi,
     \nonumber
\end{eqnarray}
and we have found that $\max |\Psi|^2=0.896$ for $\tau\le 500$.

\begin{figure}[!ht]

  \centering 
 \includegraphics[width=43mm]{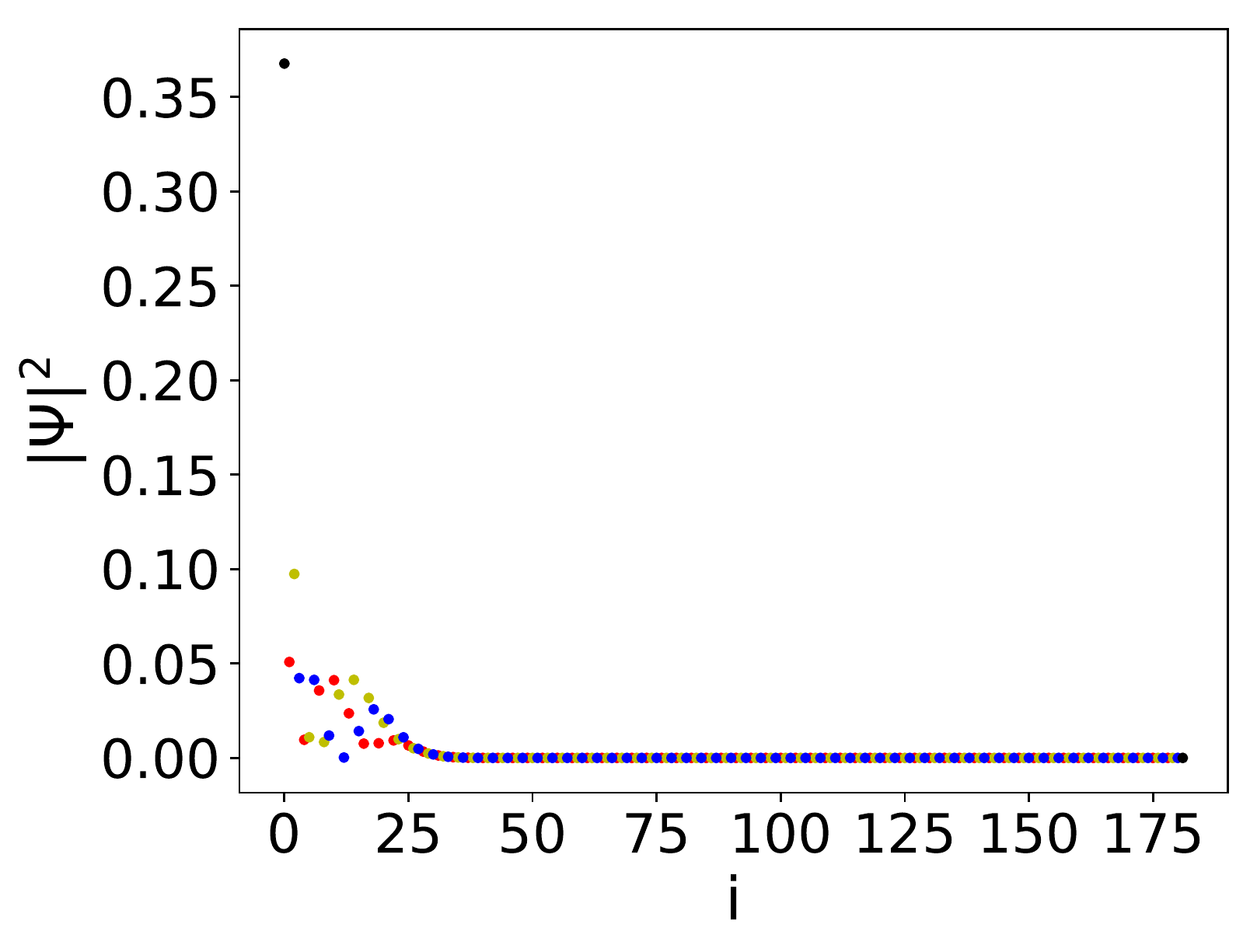}
 \includegraphics[width=43mm]{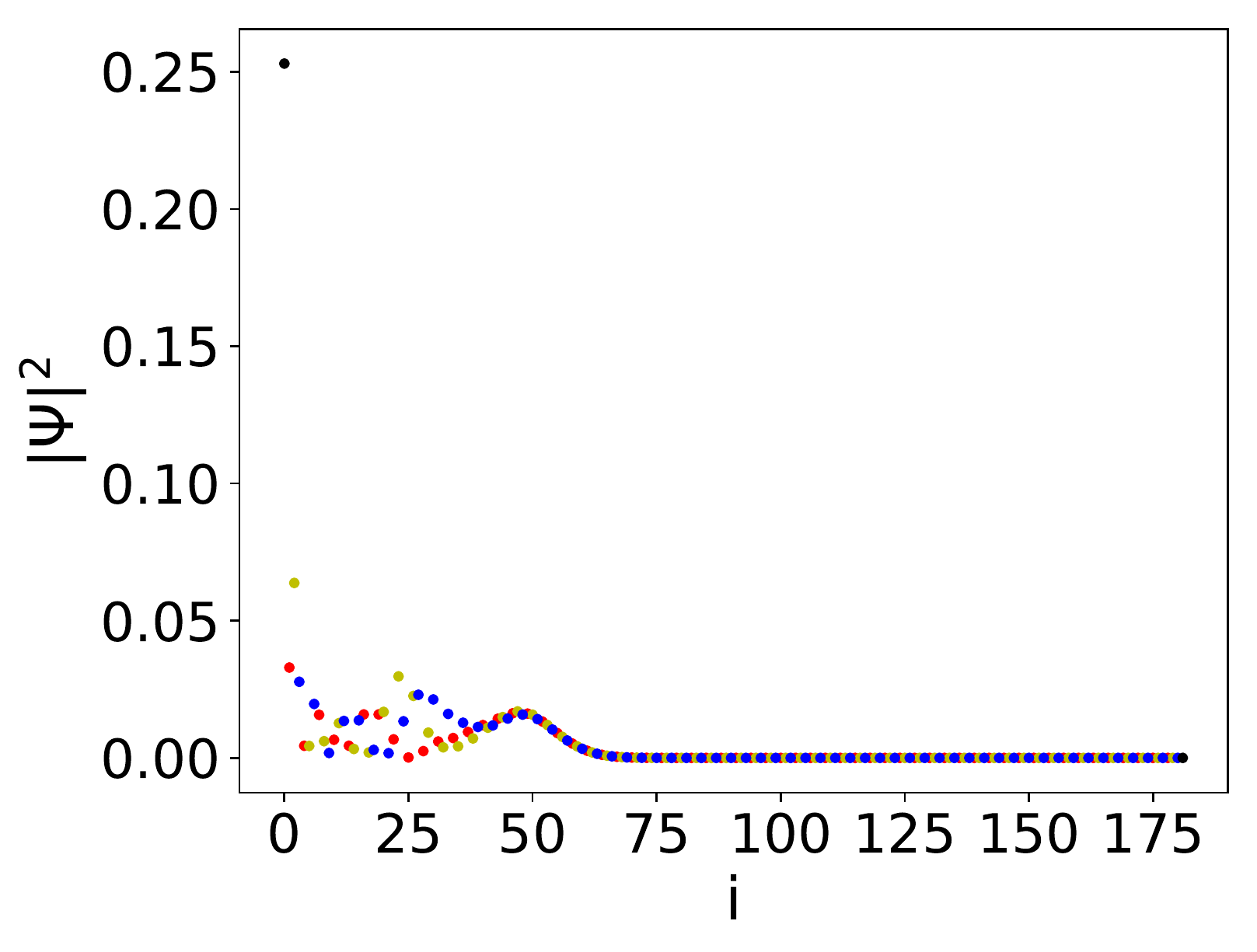}
 \includegraphics[width=43mm]{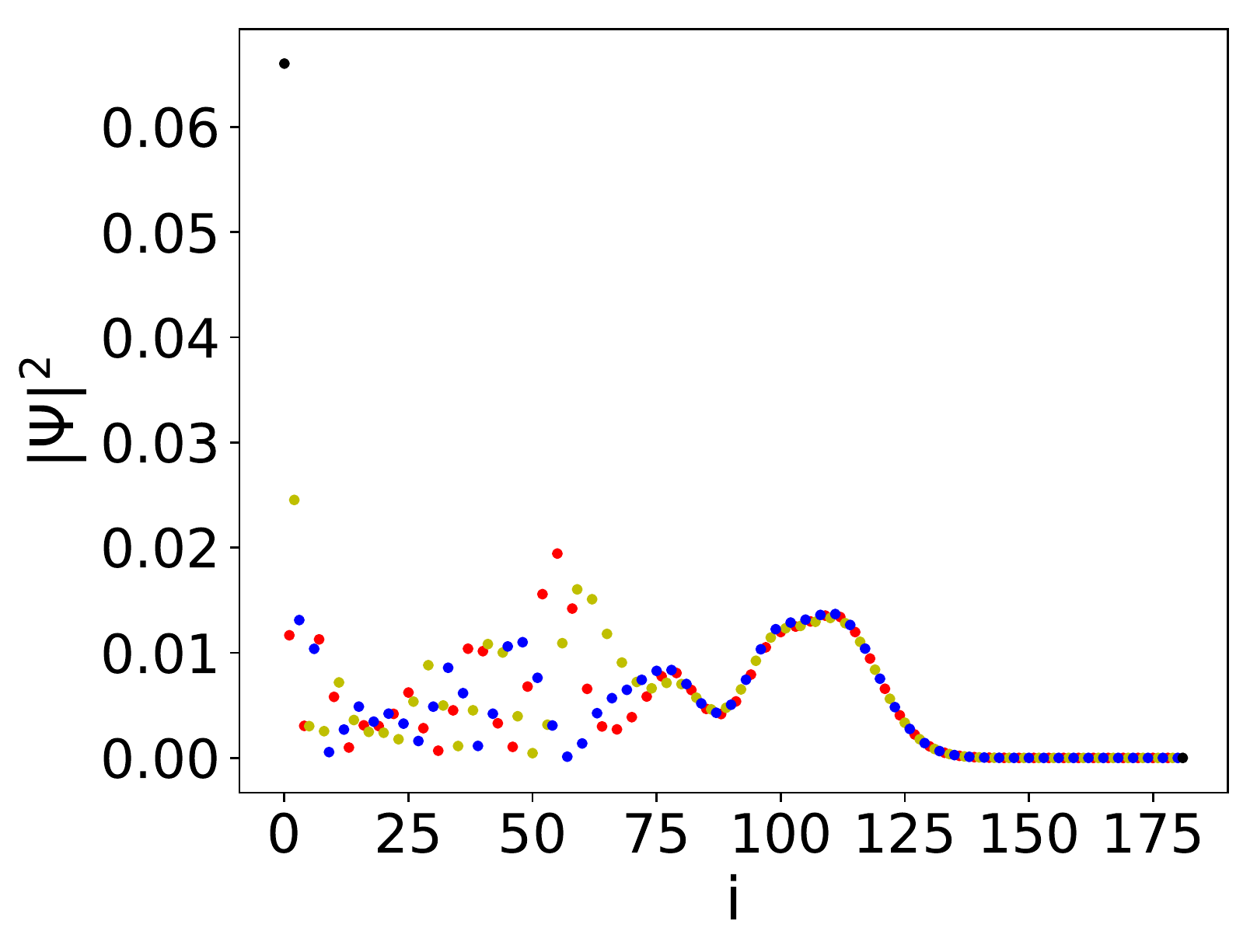}

 a \hspace{45mm} b \hspace{45mm} c
 \vspace{5mm}
 
 \includegraphics[width=43mm]{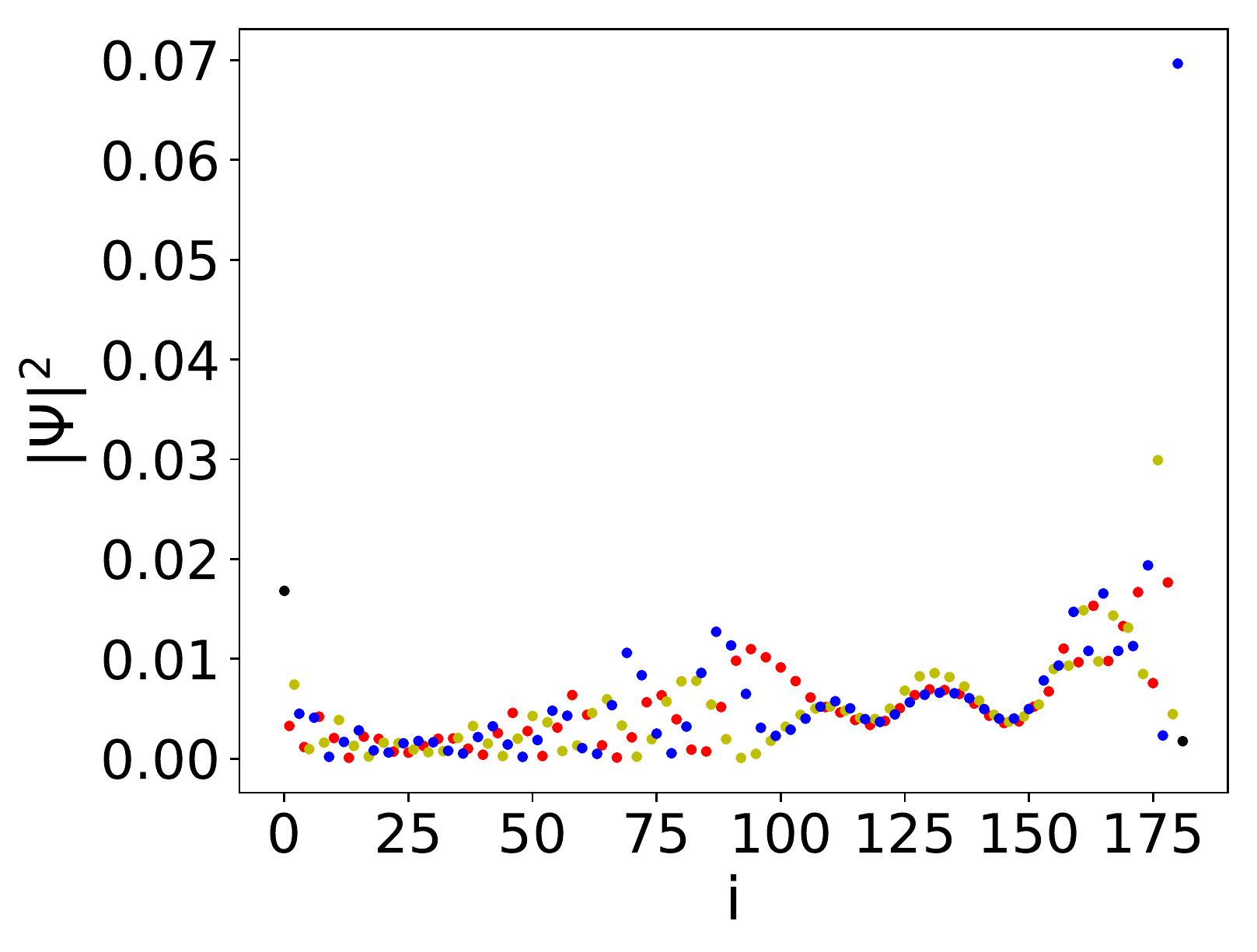}
 \includegraphics[width=43mm]{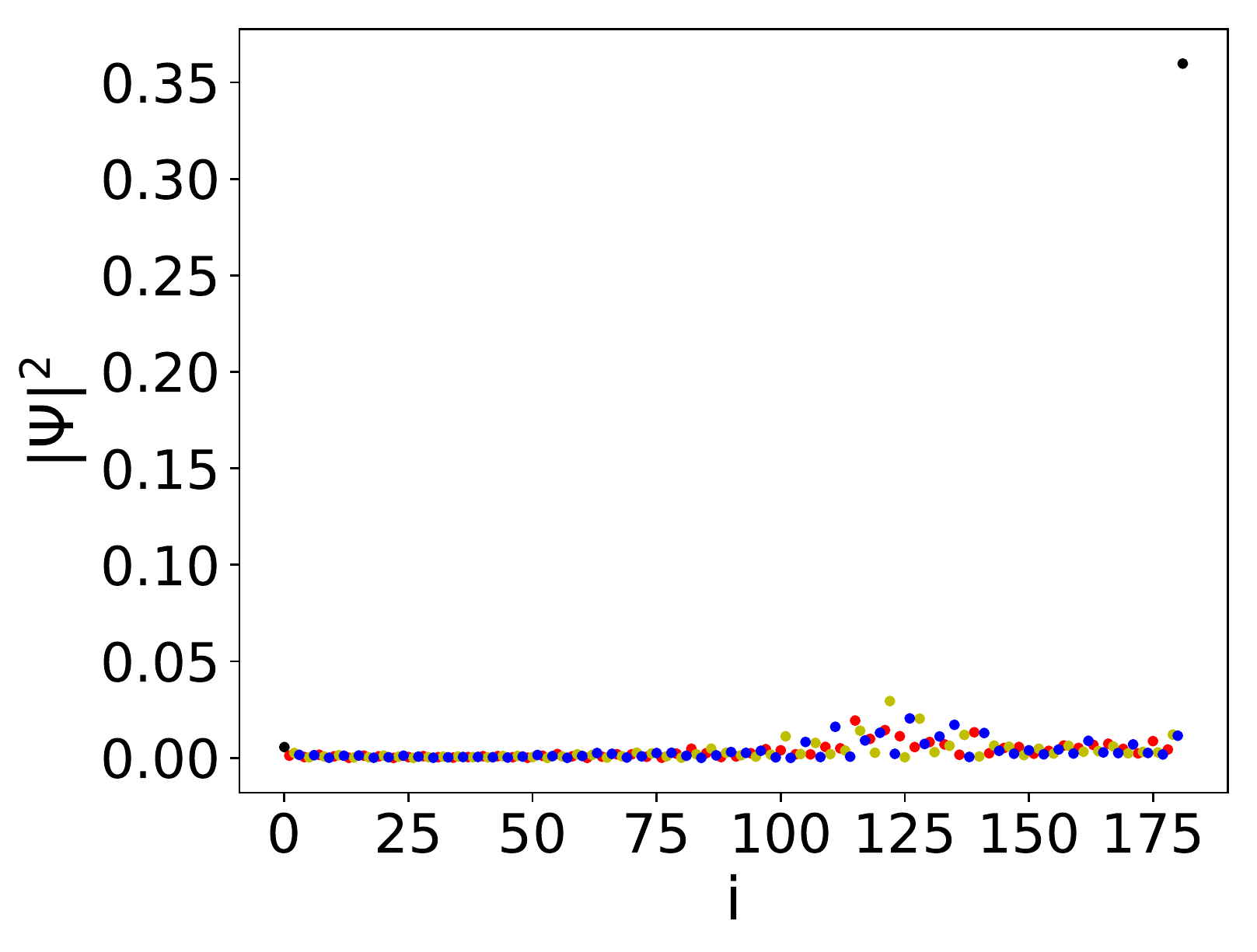}
 \includegraphics[width=43mm]{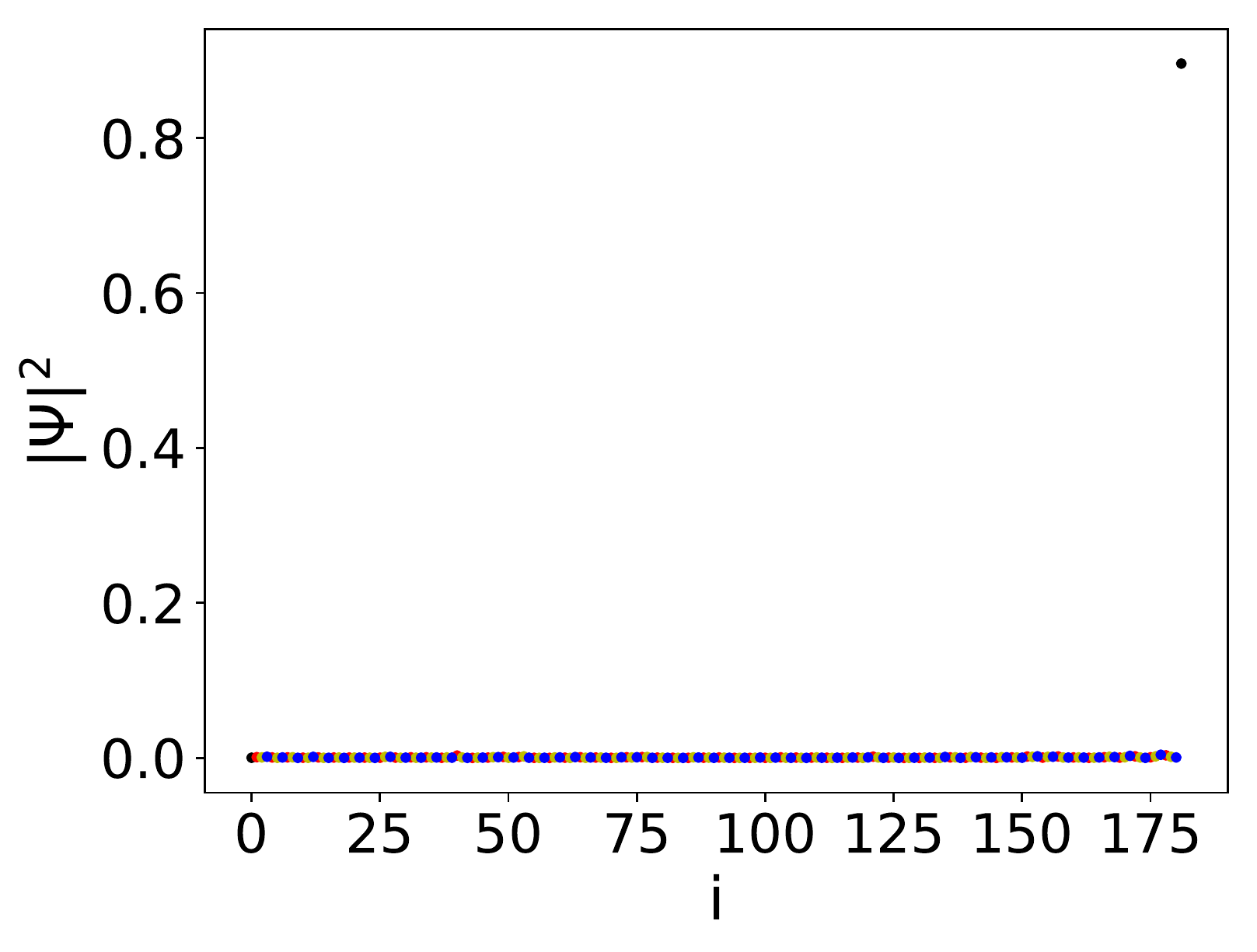}

 d \hspace{45mm} e \hspace{45mm} f
 \caption{Profile of $|\Psi|^2$ for the full homogeneous coupling during the
   transfer from donor to acceptor. a) $\tau=25$, b) $\tau=50$, c) $\tau=100$,
   d) $\tau=150$, e) $\tau=200$, f) $\tau=500$.
}
\label{fig:transfer}
\end{figure}

In Fig. \ref{fig:transfer} we present the plots of the time evolution of the electron
probability density 
for the parameters values (\ref{eq:params})
and (\ref{eq:best_hc}). We see very clearly that the electron is transferred
from
the donor onto the helix and that it then forms a localized wave that propagates
along the helix (see the movie in the supplementary material) and has a
complex structure. To understand this result, we have to recall the study of
soliton formation in alpha-helices \cite{Brizhik2004} disregarding the
donor-acceptor problem. There it has been shown that there exist several types
of solitons with different energies and symmetries.  This comes from the fact
that in the energy spectrum of alpha-helix there are 3 peptide groups per
elementary cell and so three electron energy bands which correspond to the
Davydov splitting. One of these bands is symmetric and has its minimum in the
center of the Brillouin zone, at the wavevector $k =0$, while the other two
non-symmetric lower energy bands are degenerate and have their minima at
respectively
$k_0 = \pm9\Lb/(\sqrt{3} (18 \Jb + \Lb)) $. As a result, solitons of the
first type are formed by the electron from the higher energy band and have an
energy, which is split from the higher energy band bottom.  On the other hand,
the solitons of the second type have energies which are split from the
degenerate energy band bottoms and are lower than the energy of the first type
of soliton. More importantly, there is an 'hybrid' soliton formed by the
entanglement (hybridization) of electron probabilities in the two lowest bands
due to the Jan-Teller effect and this soliton has the longest life-time. For the
alpha-helix parameter values, which we use in our simulations,
this hybrid soliton has an energy which is almost 50 times lower than the
energy of the first type of soliton.  We can expect, and indeed we will see
in what follows, that the full homogeneous coupling provide the best
conditions for launching the hybrid soliton in the helix as it has  the lowest
energy and hence leads to the highest probability for the electron to be
transported to the opposite end of the helix.

The complex soliton-like wave generated in the helix after the electron has
tunnelled into it from the donor molecule corresponds, in our numerical
simulations, to this hybrid soliton.  This hybrid soliton is not localized
on a single strand,
instead it is distributed between the strands and propagates along the helix
with some intrinsic oscillations, rather than along a particular strand,
a fact which reflects its hybrid nature. The propagation of this localized
polaron is followed by what looks like incoherent ripples. These ripples
describe the radiated sound waves in the helix. This is because our system
is not completely integrable and while most of the initial electron energy
is transferred to the soliton, some of it is converted into oscillating
soliton 'tails'.

\begin{figure}[!ht]
    \centering 
 \includegraphics[width=10cm]{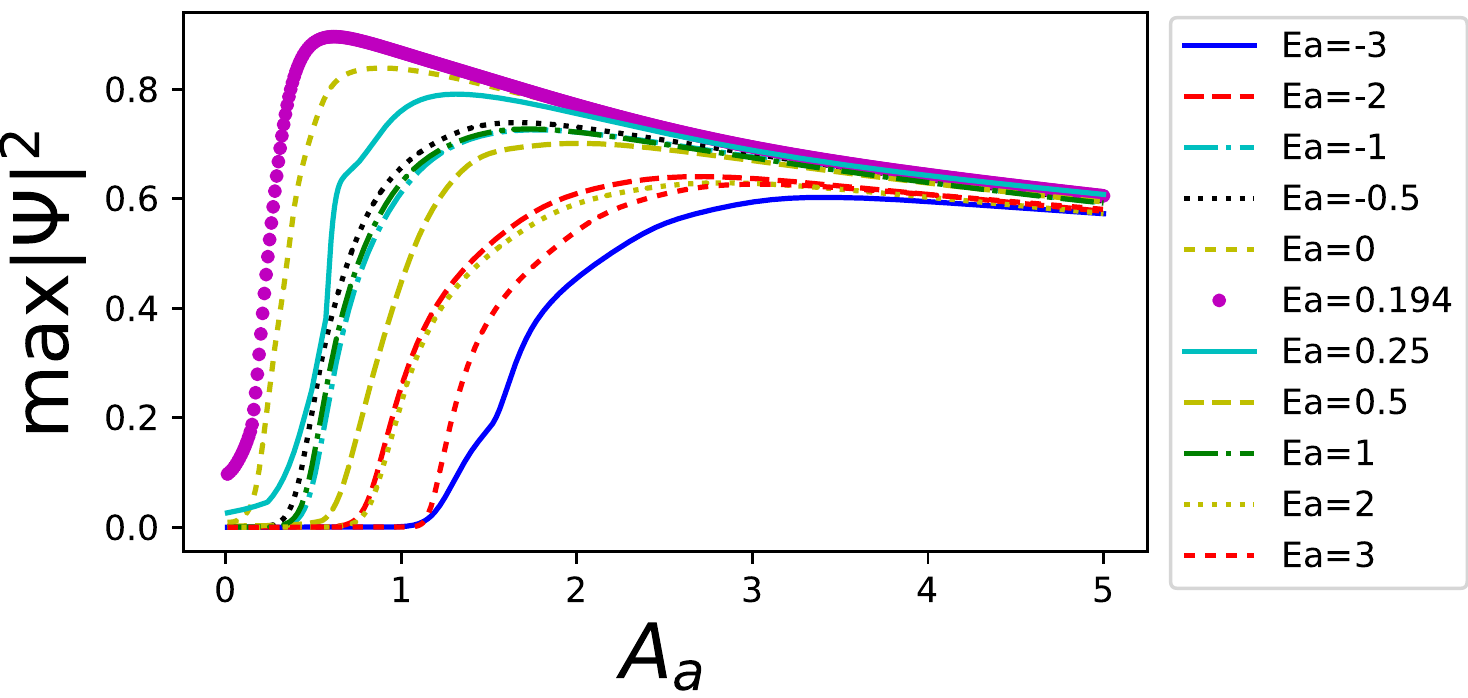}
 \caption{Full homogeneous coupling. The plot of $\max(|\Psi_{N+1}|^2)$ for
   $\tau \le 500$ as a function of $A_{a}$ for different values
   of $\E_a$ and the parameters values (\ref{eq:best_hc}).
}
\label{fig:FF_Aa_Ea0}
\end{figure}

\begin{figure}[!ht]
  \centering 
  \includegraphics[width=10cm]{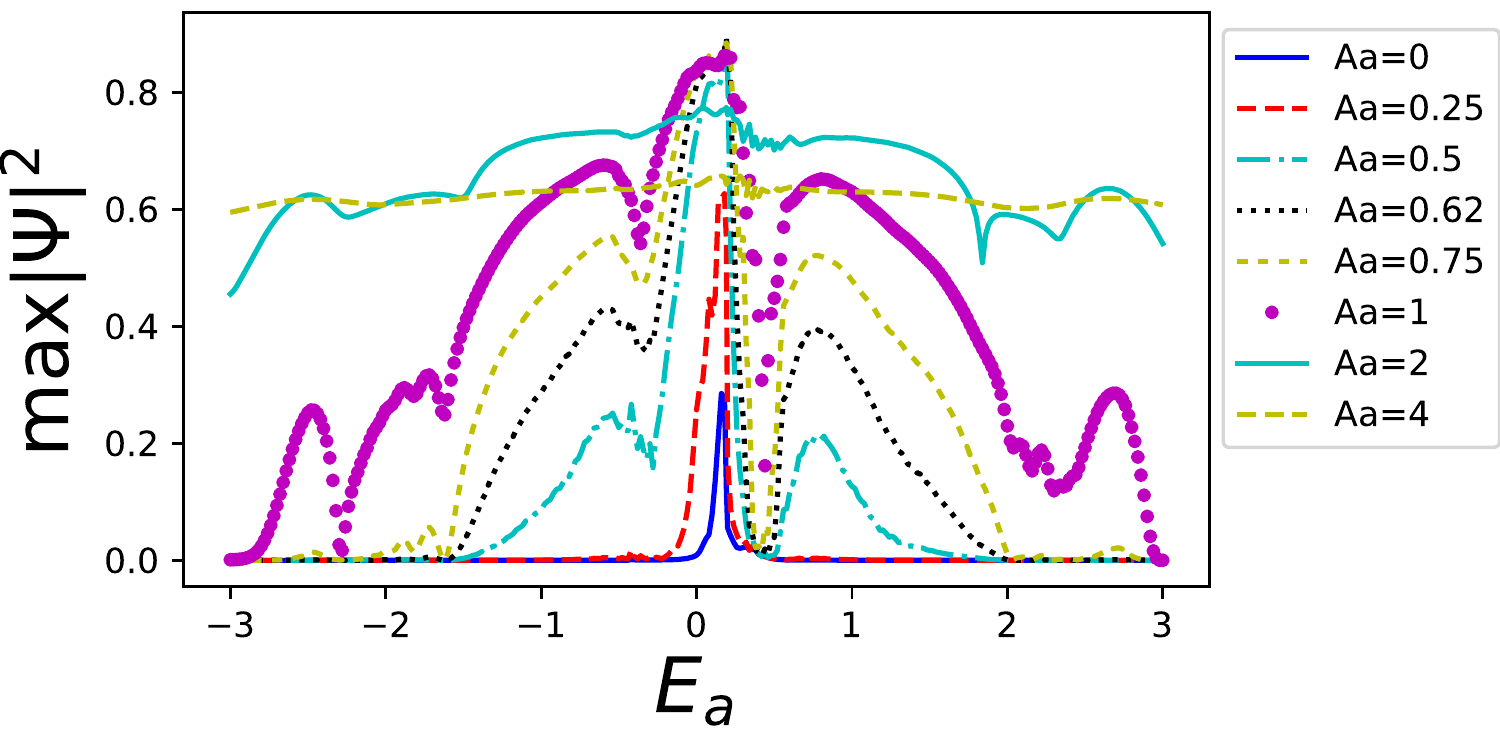}
  \caption{Full homogeneous coupling. The plot of $\max(|\Psi_{N+1}|^2)$ for
    $\tau \le 500$ as a function of $E_{a}$ for different values
   of $A_{a,1}=A_{a,2}=A_{a,3}=A_a$ and the parameters values (\ref{eq:best_hc}).
}
\label{fig:FF_Ea0_Aa}
\end{figure}

\begin{figure}[!ht]
    \centering 
 \includegraphics[width=10cm]{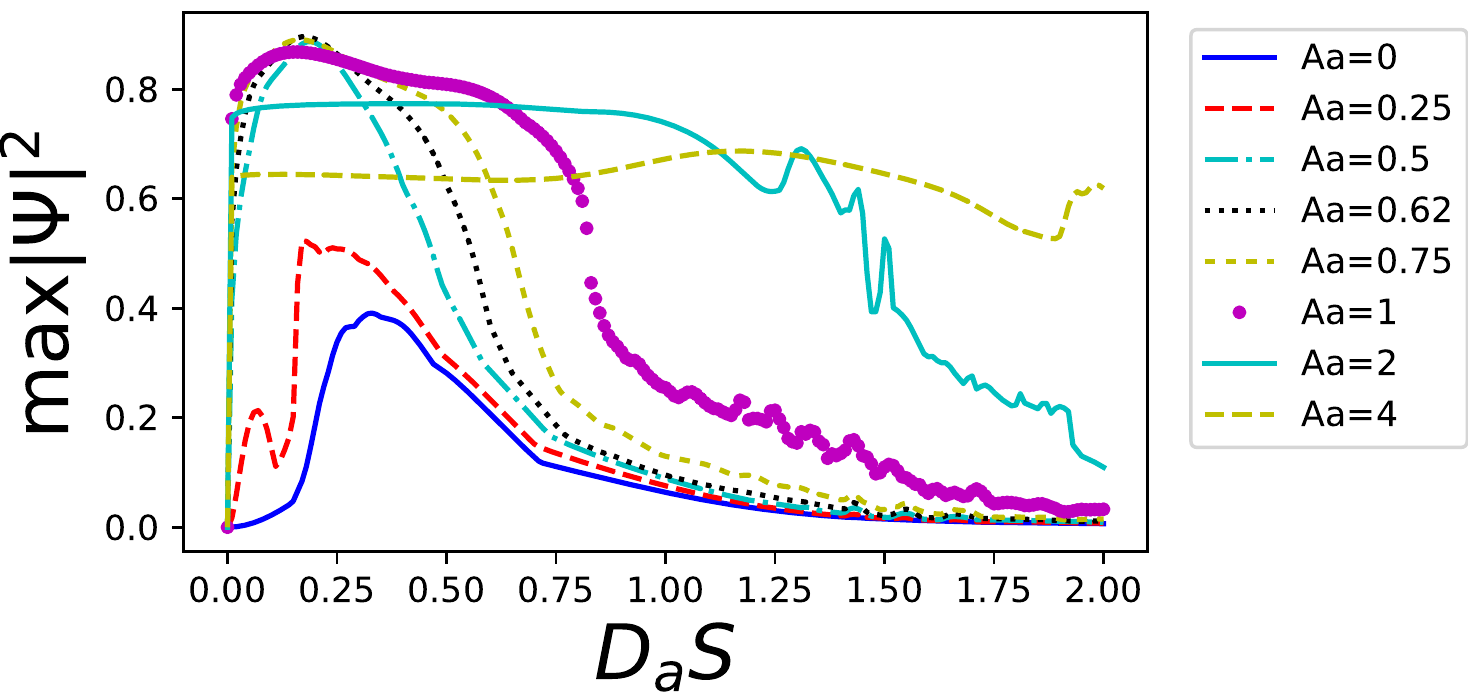}
 \caption{Full homogeneous coupling. The plot of $\max(|\Psi_{N+1}|^2)$ for
   $\tau \le 500$
   as a function of $D_{a}S=D_{a,\ell}/J$ for different values
   of $A_{a,1}=A_{a,2}=A_{a,3}=A_a$ and the parameters values (\ref{eq:best_hc}).
}
\label{fig:FF_Das_Aa}
\end{figure}

\begin{figure}[!ht]
    \centering 
    \includegraphics[width=10cm]{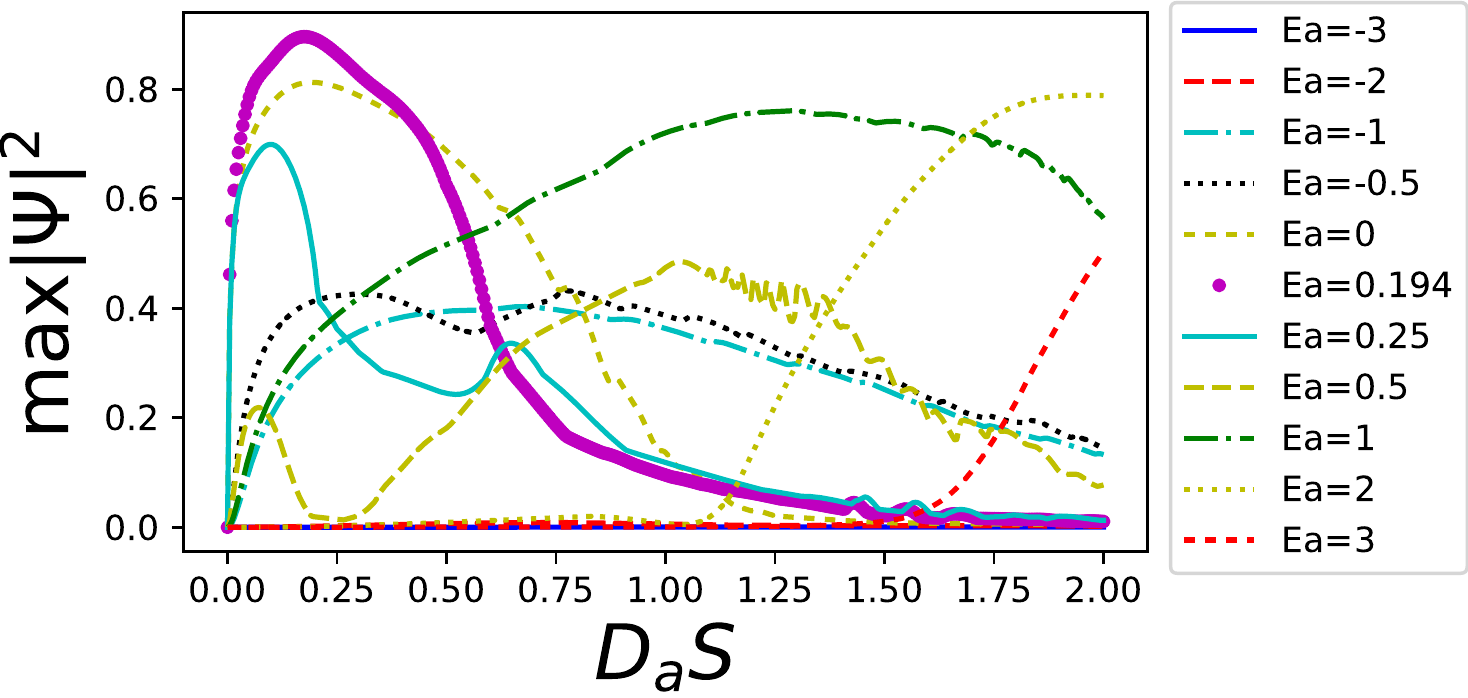}
    \caption{Full homogeneous coupling. The plot of $\max(|\Psi_{N+1}|^2)$ for
      $\tau \le 500$
   as a function of $D_{a}S=D_{a,\ell}/J$ for different values
   of $\E_a$ and the parameters values (\ref{eq:best_hc}).
}
\label{fig:FF_Das_Ea0}
\end{figure}

\begin{figure}[!ht]
    \centering 
 \includegraphics[width=10cm]{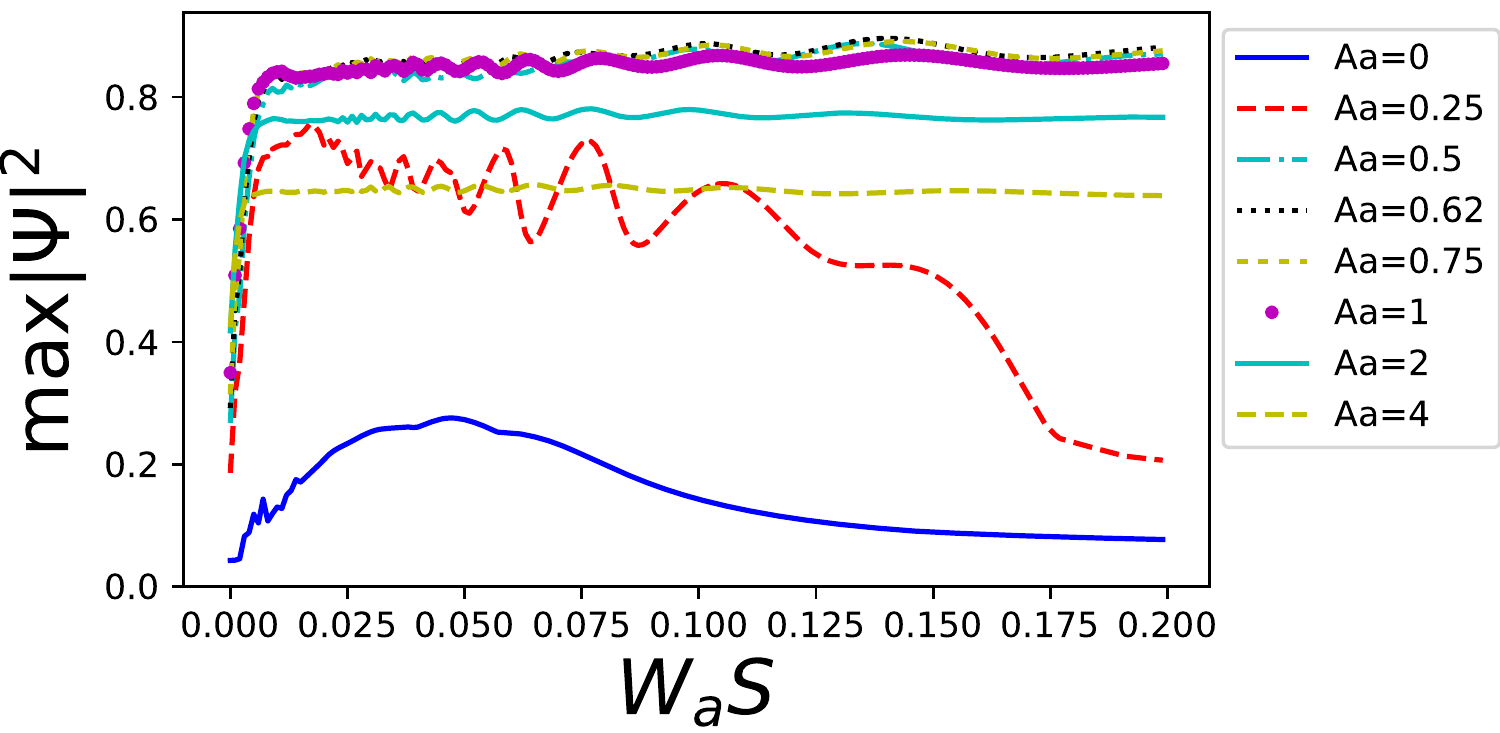}
 \caption{Full homogeneous coupling. The plot of $\max(|\Psi_{N+1}|^2)$ for
   $\tau \le 500$
   as a function of $W_{a}S=W_{a,\ell}/W$ for different values
   of $A_{a,1}=A_{a,2}=A_{a,3}=A_a$ and the parameters values (\ref{eq:best_hc}).
}
\label{fig:FF_WaS_Aa}
\end{figure}

\begin{figure}[!ht]
    \centering 
 \includegraphics[width=10cm]{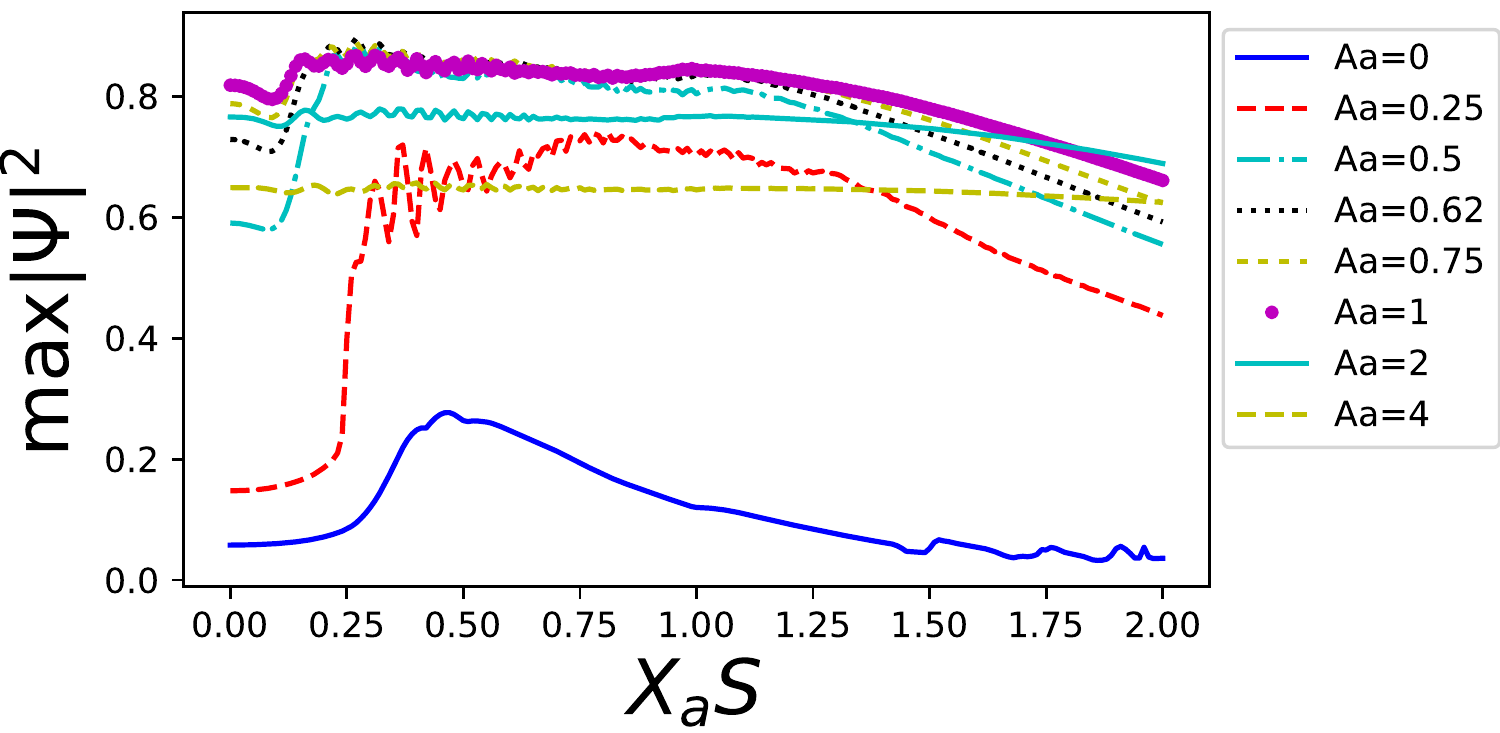}
 \caption{Full homogeneous coupling. The plot of $\max(|\Psi|^2)$ for
   $\tau \le 500$
   as a function of $\chi_{a}S=\chi_{a,\ell}/\chi$ for different
   values
   of $A_{a,1}=A_{a,2}=A_{a,3}=A_a$ and the parameters values (\ref{eq:best_hc}).
}
\label{fig:FF_XaS_Aa}
\end{figure}

We have then studied how $\max |\Psi_{N+1}|^2$ varies when the acceptor
parameters are varied around their optimal value.
This is shown in Figs. 
\ref{fig:FF_Aa_Ea0}-\ref{fig:FF_XaS_Aa}.

To perform these simulations, we have defined the following parameters
\begin{eqnarray}
D_{a}S&=&\frac{D_{a,\ell}}{J},\qquad   
W_{a}S=\frac{W_{a,\ell}}{W},\qquad   
X_{a}S=\frac{\chi_{a,\ell}}{\chi},
\label{eq:FF_param} 
\end{eqnarray}
which relate the different parameters of the donor and the acceptor to the
corresponding ones on the peptide chain.

From Figs. \ref{fig:FF_Aa_Ea0} and \ref{fig:FF_Ea0_Aa}, we first note that
the value of the acceptor electron energy $E_a$ has to be
relatively small for the electron to be transferred to the acceptor, and that
the values of $\E_a$ and $A_a$ must be finely tuned for a good 'capture' of
the electron.
The parameters $D_a$ and $W_a$ and $\chi_a$, on the other hand, offer a much
broader tolerance when $\E_a$ and $A_a$ are correctly tuned (see Figs.
\ref{fig:FF_Das_Aa} to \ref{fig:FF_XaS_Aa}). This result has a clear physical
meaning, since at this last stage of the electron transport the dominant
parameters are the strength of the exchange interaction of the acceptor with
the helix and the value of the on-site energy level on the acceptor, while
on the other hand, the electron-lattice coupling and the elasticity of the
acceptor-helix bond are much less important. This being said, the last stage
of the transport process is only possible if a proper soliton has been
launched on the helix, carrying most of the initial energy and
electron probability to the acceptor with minimal energy dissipation into
the lattice vibrations and heat generation.

As we will see in the next subsections, the effectiveness of the soliton
formation and its parameters are determined by (i) the helix parameters,
mainly by the electron-lattice coupling and strand elasticity, and (ii) by
the number of helix strands coupled to the donnor.

\subsection{Single Strand Coupling}
In this section we couple the donor only to the first node of the chain:
$D_{d,2}=D_{d,3}=W_{d,2}=W_{d,3}=\chi_{d,2}=\chi_{d,3}=0$.
We obtain the best transfer from the donor to the chain for the following
donor parameters:
\begin{eqnarray}
  &&\E_d=0.25,\quad D_{d,1}=0.38\,J, \quad
     W_{d,1}=0.32\,W,\quad  \chi_{d,1}=0.62\,\chi.
     \label{eq:best_sc}
\end{eqnarray}

Such conditions for creation of the soliton in the helix are not the
optimal ones, since in the soliton formation there will be two contradicting
tendencies:  redistribution of the electron between the three peptide groups
within the same unit cell and its dispersion to the nearest unit cell.
These processes will be accompanied by energy dissipation much stronger than
in the case of the full homogeneous coupling, and thus will result in the
generation of a much weaker soliton i.e., in a less efficient transport of the
electron along the helix. 

Similarly, the type of coupling between the helix and the acceptor plays an
important role in the electron transport, as we will see from this and the
next sub-section. First, we consider the coupling of the acceptor to  the same
strand as the one to which the donor is coupled (single-strand coupling),
setting 
$A_{a,2}=A_{a,3}=D_{a,2}=D_{a,3}=W_{a,2}=W_{a,3}=\chi_{a,2}=\chi_{a,3}=0$.

We found that the best parameters to obtain a transfer of the electron
to the acceptor are
\begin{eqnarray}
  &&\E_d=0.25,\quad D_{d,1}=0.38\,J, \quad
     W_{d,1}=0.32\,W,\quad  \chi_{d,1}=0.62\,\chi,
  \\
  &&A_{a,1}=6.5, \quad \E_a=0.265,\quad D_{a,1}=0.3\,J, \quad
     W_{a,1}=0.37\,W,\quad  \chi_{a,1}=\chi.\nonumber
     \label{eq:best_ssc}
\end{eqnarray}

As one could expect, the maximum value of the electron probability on the acceptor $|\Psi_{N+1}|^2$ for $\tau \le 500$ in this case is much lower, than in  the full homogeneous case,   and is equal to 
 $0.21839$, only, showing
that in this configuration, the electron is only transferred to 
acceptor with a 20\% probability. As this is quite small we did not study the
variation of $\max |\Psi_{N+1}|^2$ around these optimal values of the parameters.

\subsection{End-to-End Coupling}
To couple the acceptor to the last peptide of the helix, we take
$A_{a,1}=A_{a,2}=D_{a,1}=D_{a,2}=W_{a,1}=W_{a,2}=\chi_{a,1}=\chi_{a,2}=0$.
In this case we have obtained the best transfer using the following parameters:
\begin{eqnarray}
  &&\E_d=0.25,\quad D_{d,1}=0.38\,J, \quad
     W_{d,1}=0.32\,W,\quad  \chi_{d,1}=0.62\,\chi,\\
  &&A_{a,3}=1.98, \quad \E_a=0.276,\quad D_{a,3}=0.29\,J, \quad
     W_{a,3}=0.002\,W,\quad  \chi_{a,3}=0.04\,\chi,\nonumber
     \label{eq:best_eec}
\end{eqnarray}
and, with this choice, we have found that $\max |\Psi_{N+1}|^2=0.642558$, which is higher than in the previous case, although lower than in the case of the full homogeneous coupling.

\begin{figure}[!ht]
    \centering 
 \includegraphics[width=10cm]{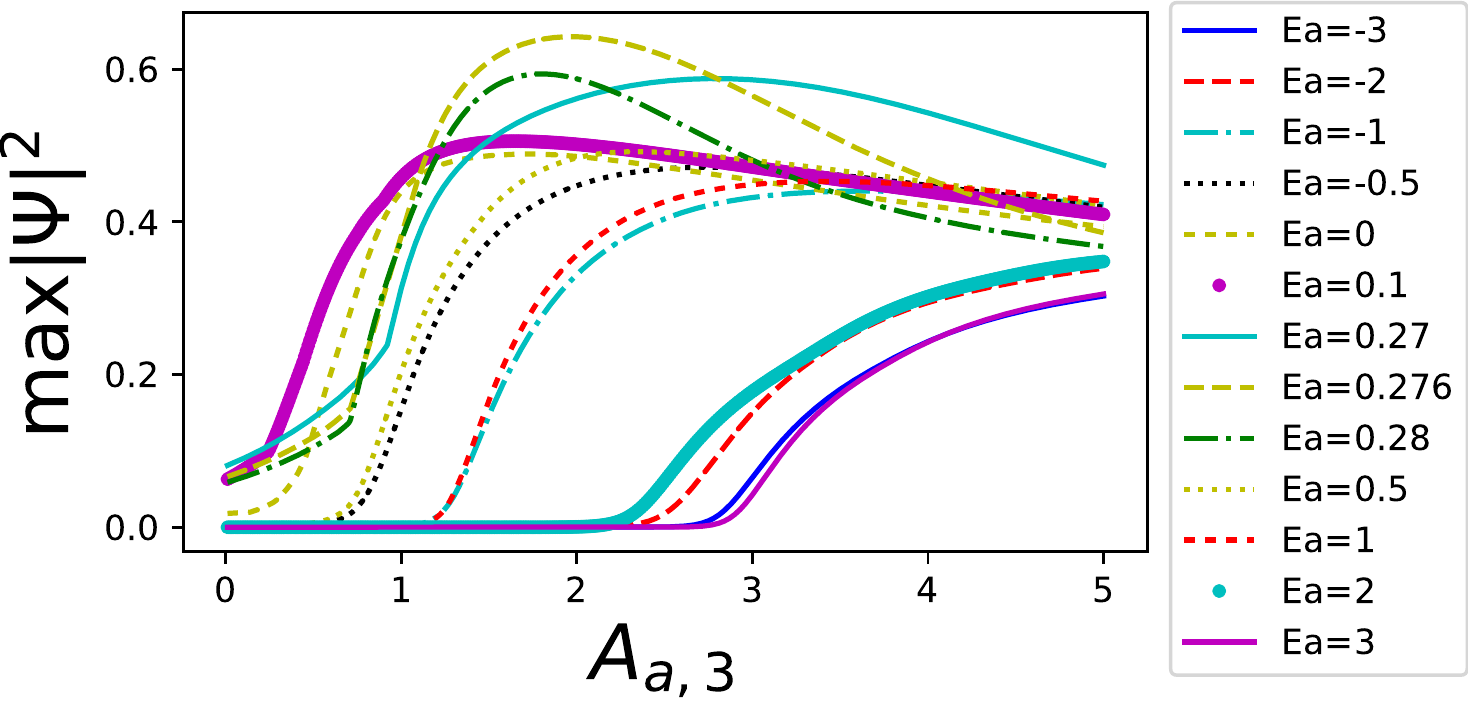}
 \caption{End-to-end coupling. The plot of $\max |\Psi_{N+1}|^2$ for
   $\tau \le 500$ as a function of $A_{a,3}$ for different values
   of $\E_a$ and the parameters values (\ref{eq:best_eec}).
}
\label{fig:EE_Aa_Ea0}
\end{figure}

\begin{figure}[!ht]
    \centering 
 \includegraphics[width=10cm]{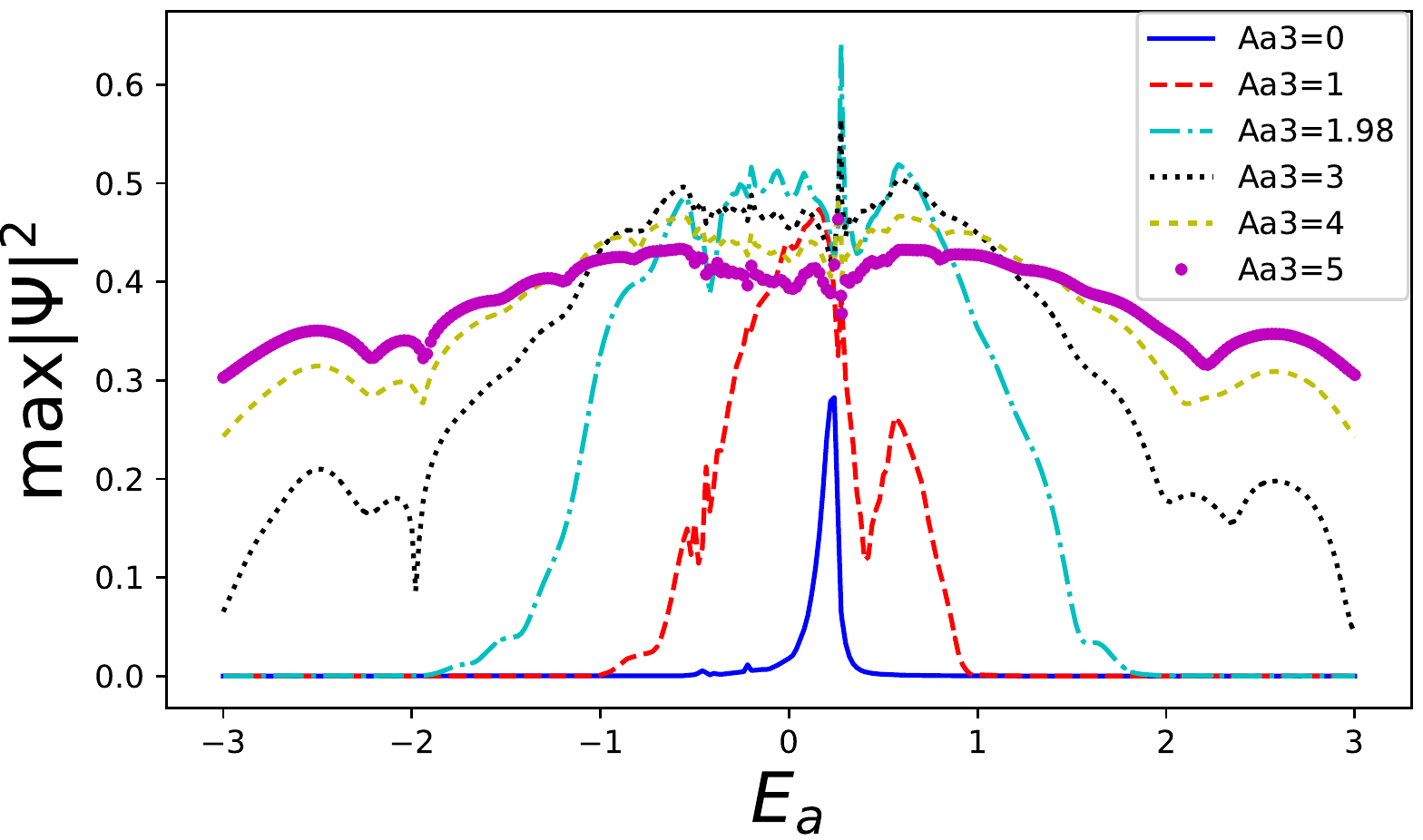}
 \caption{End-to-end coupling.  The plot of $\max |\Psi_{N+1}|^2$ for
   $\tau \le 500$ as a function of $E_a$ for different values
   of $A_{a,3}$ and the parameters values (\ref{eq:best_eec}).
}
\label{fig:EE_Ea0_Aa}
\end{figure}

\begin{figure}[!ht]
    \centering 
 \includegraphics[width=10cm]{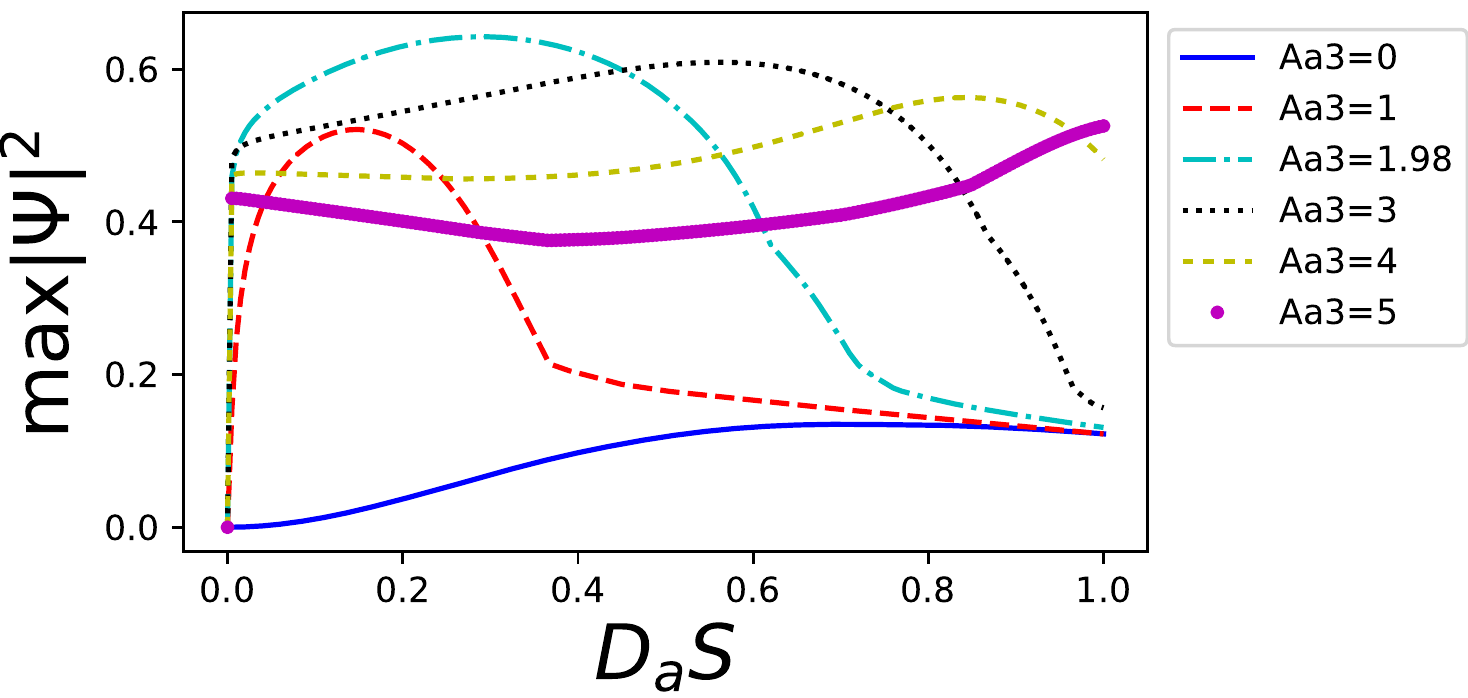}
 \caption{End-to-end coupling. The plot of $\max |\Psi_{N+1}|^2$ for
   $\tau \le 500$ as a function of $D_{a}S=D_{a,\ell}/J$ for different values
   of $=A_{a,3}$ and the parameters values (\ref{eq:best_eec}).
}
\label{fig:EE_Das_Aa}
\end{figure}

\begin{figure}[!ht]
    \centering 
 \includegraphics[width=10cm]{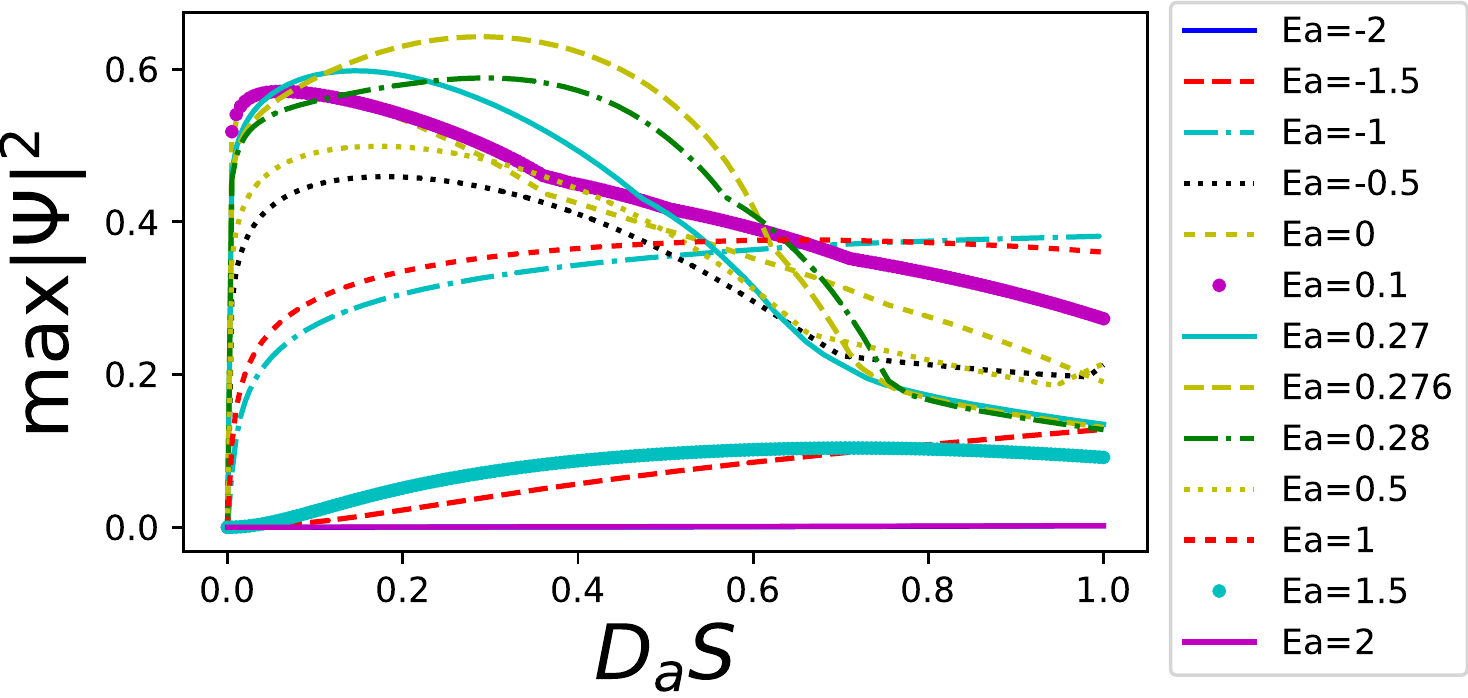}
 \caption{End-to-end coupling. The plot of $\max |\Psi_{N+1}|^2$  for
   $\tau \le 500$ as a function of $D_{a}S=D_{a,\ell}/J$ for different values
   of $\E_a$ and the parameters values (\ref{eq:best_eec}).
}
\label{fig:EE_Das_Ea0}
\end{figure}

\begin{figure}[!ht]
    \centering 
 \includegraphics[width=10cm]{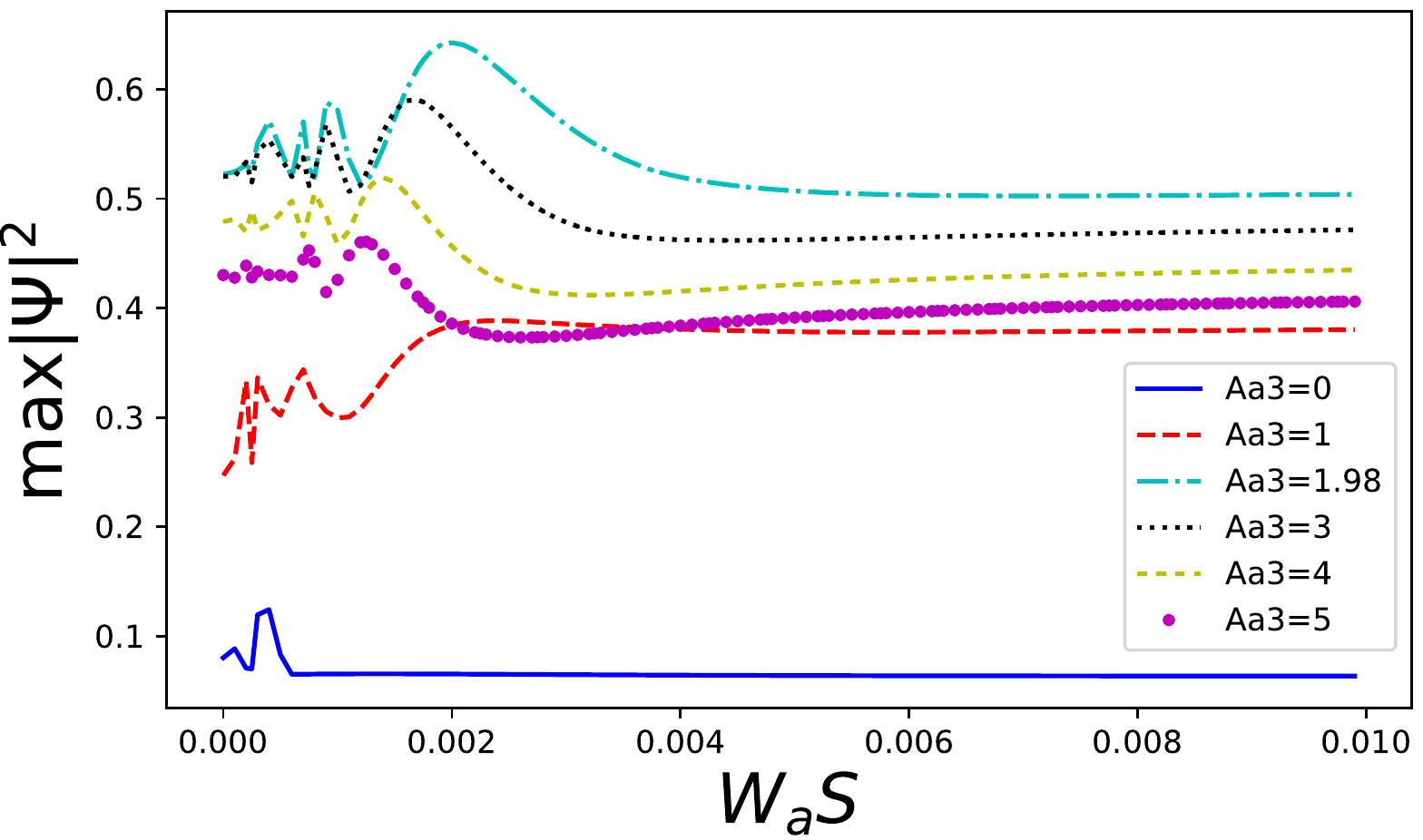}
 \caption{End-to-end coupling. The plot of $\max |\Psi_{N+1}|^2$ for
   $\tau \le 500$ as a function of $W_{a}S=W_{a,\ell}/W$ for different values
   of $A_{a,3}$ and the parameters values (\ref{eq:best_eec}).
}
\label{fig:EE_WaS_Aa}
\end{figure}

\begin{figure}[!ht]
    \centering 
 \includegraphics[width=10cm]{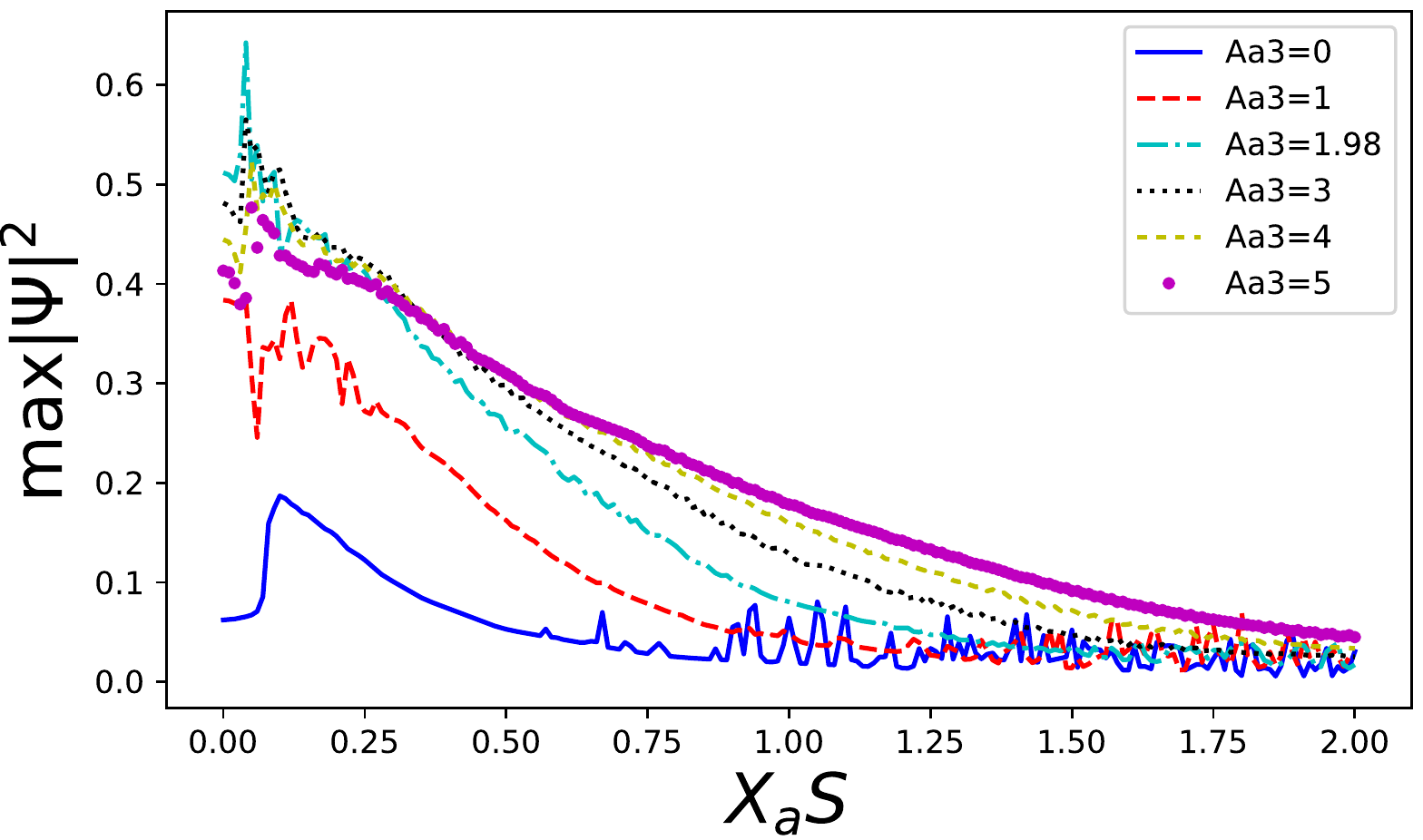}
 \caption{End-to-end coupling. The plot of $\max|\Psi_{N+1}|^2$ for
   $\tau \le 500$ 
   as a function of $\chi_{a}S=\chi_{a,\ell}/\chi$ for different values
   of $A_{a,3}$ and the parameters values (\ref{eq:best_eec}).
}
\label{fig:EE_XaS_Aa}
\end{figure}

We have then studied how $\max |\Psi_{N+1}|^2$ varies when the acceptor
parameters are varied around their optimal value. This is shown in Figs.
\ref{fig:EE_Aa_Ea0} to \ref{fig:EE_XaS_Aa}.

As with the full homogeneous coupling, we have found that the absorption is
mainly controlled by a fine tuning between $A_{a,3}$ and $E_a$ but that
there is a broader tolerance for the values of $D_a$, $W_a$ and $\chi_a$.

Having analysed the parameter stability of our model we now turn to the
study of its thermal stability.

\section{Thermal Stability of the Soliton Mediated Electron Transport}
So far, in the study of our model, we have not taken into account any thermal fluctuations. To include them we have modified the model by
 adding the following Langevin terms to the
equations for $U_n$:
\begin{eqnarray}
  L_n = F_n(\tau)-\Gamma \frac{du_n}{d\tau} ,
\end{eqnarray}
where $\Gamma$ is an absorption parameter and $F_n(\tau)$ represents the thermal
noise modelled as a Gaussian white noise of zero mean value and variance given
by
\begin{eqnarray}
  \langle F_n(\tau_1)\,F_m(\tau_2) \rangle =
    2 \Gamma k T \delta(\tau_1-\tau_2)\delta_{n,m},
\end{eqnarray}
where, $k$ is Boltzmann coefficient, and for the dimensional thermal energy
$k \overline {T}$, we have $ kT= k \overline{T}/\hbar \nu$.
To implement this numerically, $F(\tau)$ has to be kept constant during each
time step $d\tau$ and so we have used $\delta(\tau_1-\tau_2)=1/d\tau$.

For each temperature, we have performed $100$ simulations and computed the
mean values of $\max|\Psi_a|$, for $\tau\le 100$, obtained from these
simulations.

At physiological temperature, $\overline{kT} \approx 0.025 eV$ which in our
adimensional units corresponds to $7.12$. We have thus varied $kT$ between $0$
and $10$ to capture the physiological conditions when $\Jb$ is smaller than
$0.035 eV$.

\begin{figure}[!ht]
  \vspace{-1mm}
    \centering 
 \includegraphics[width=10cm]{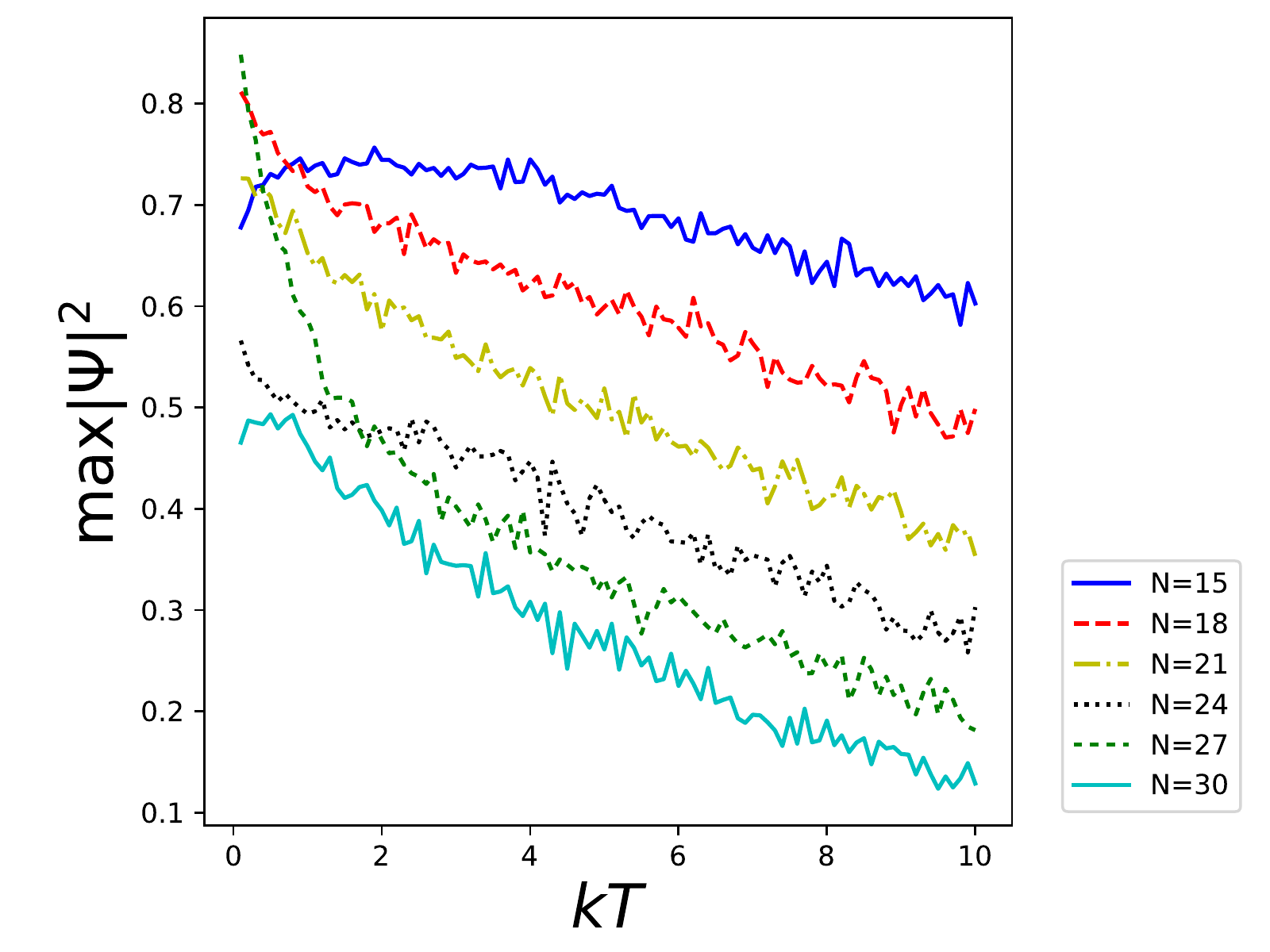}
  \vspace{-1mm}
  \caption{Full homogeneous coupling. The plot of $\max |\Psi_{N+1}|^2$ for
    $\tau \le 100$ as a function of $kT$ for different values of the chain
    length $N$. $\Gamma=0.2$.
}
\label{fig:FF_kt}
\end{figure}

\begin{figure}[!ht]
  \vspace{-1mm}
\centerline{
 \includegraphics[width=80mm]{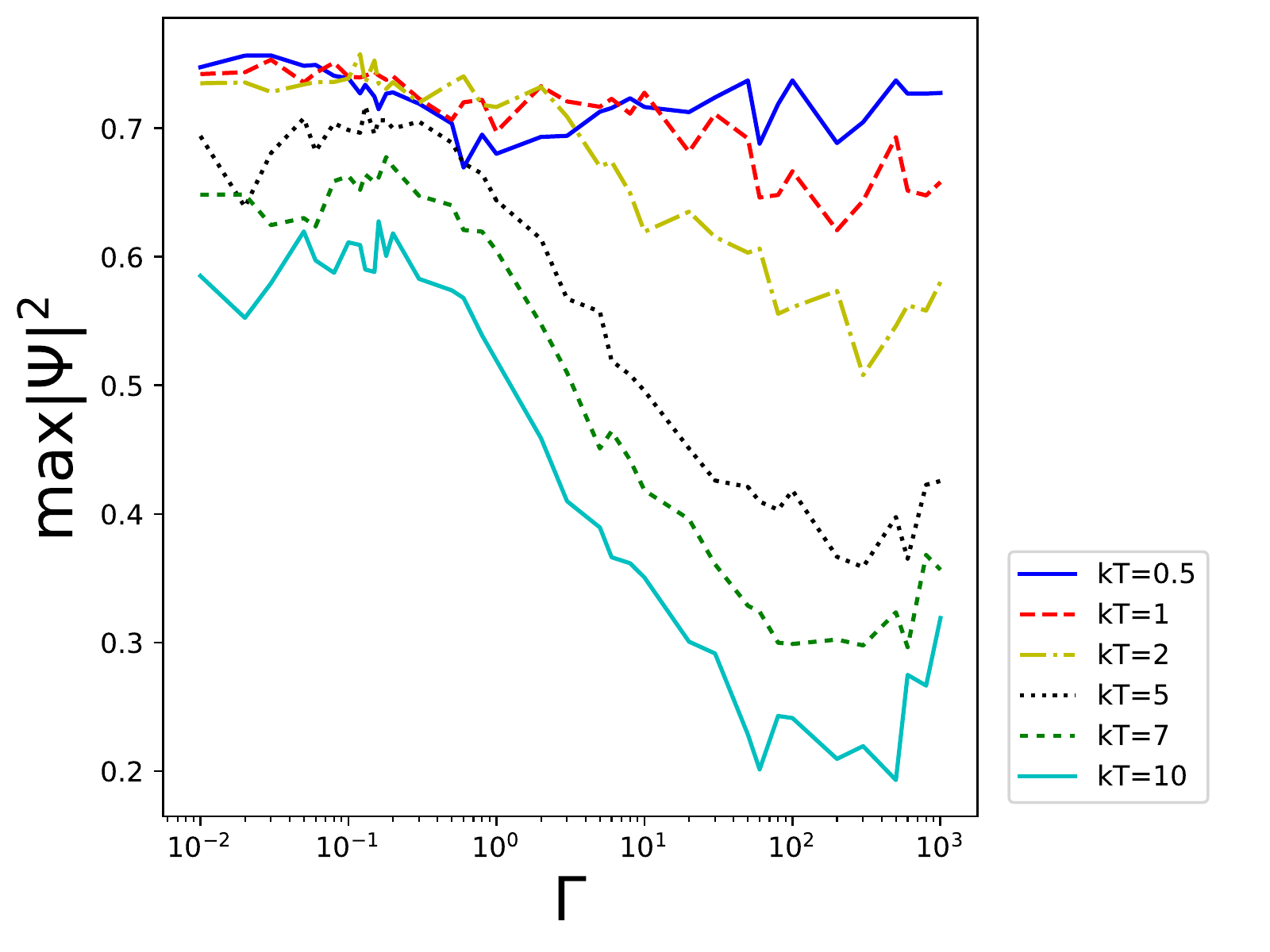}
 \includegraphics[width=80mm]{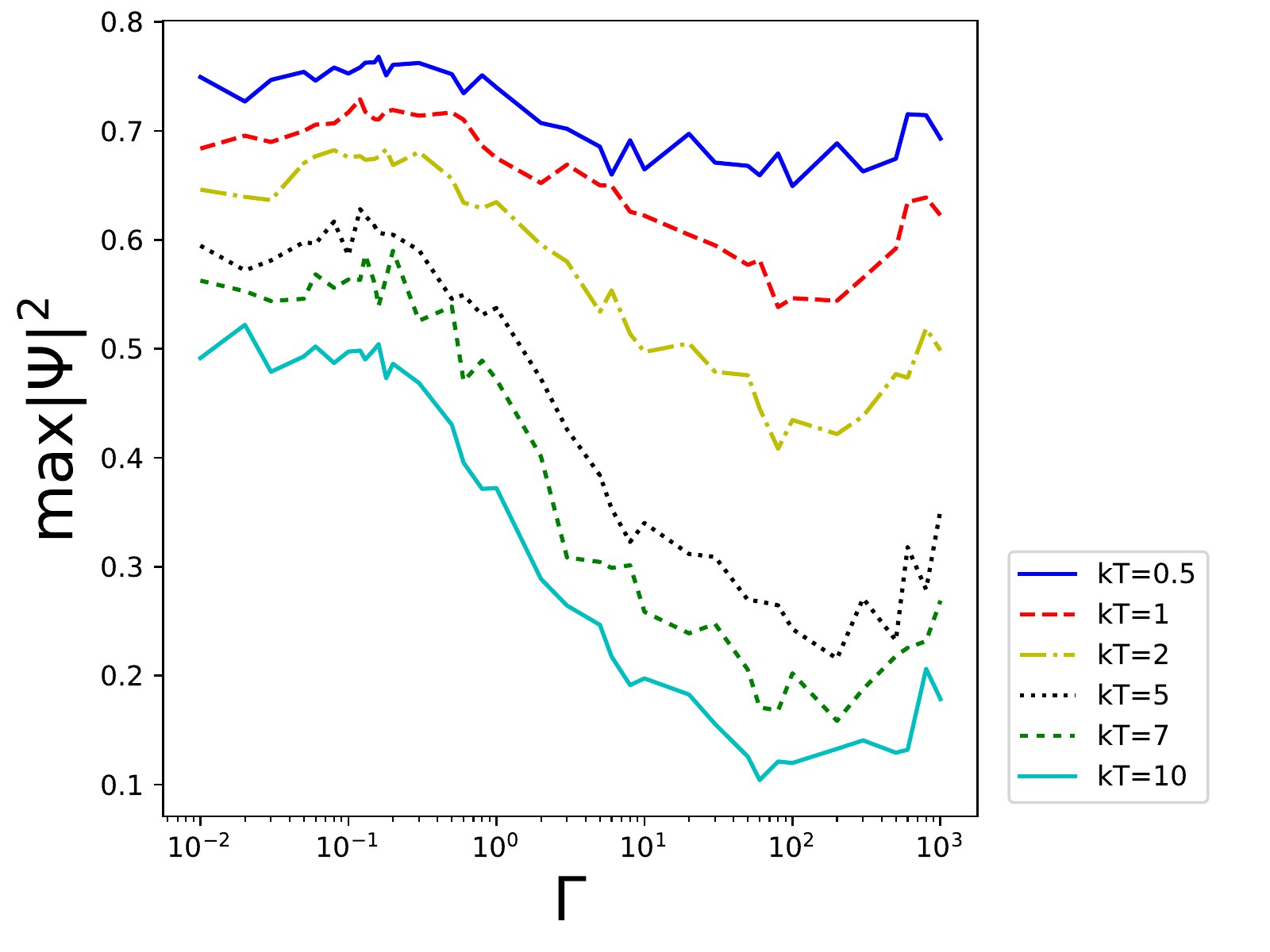}
 }
\centerline{ a \hspace{80mm} b }
 \centerline{
 \includegraphics[width=80mm]{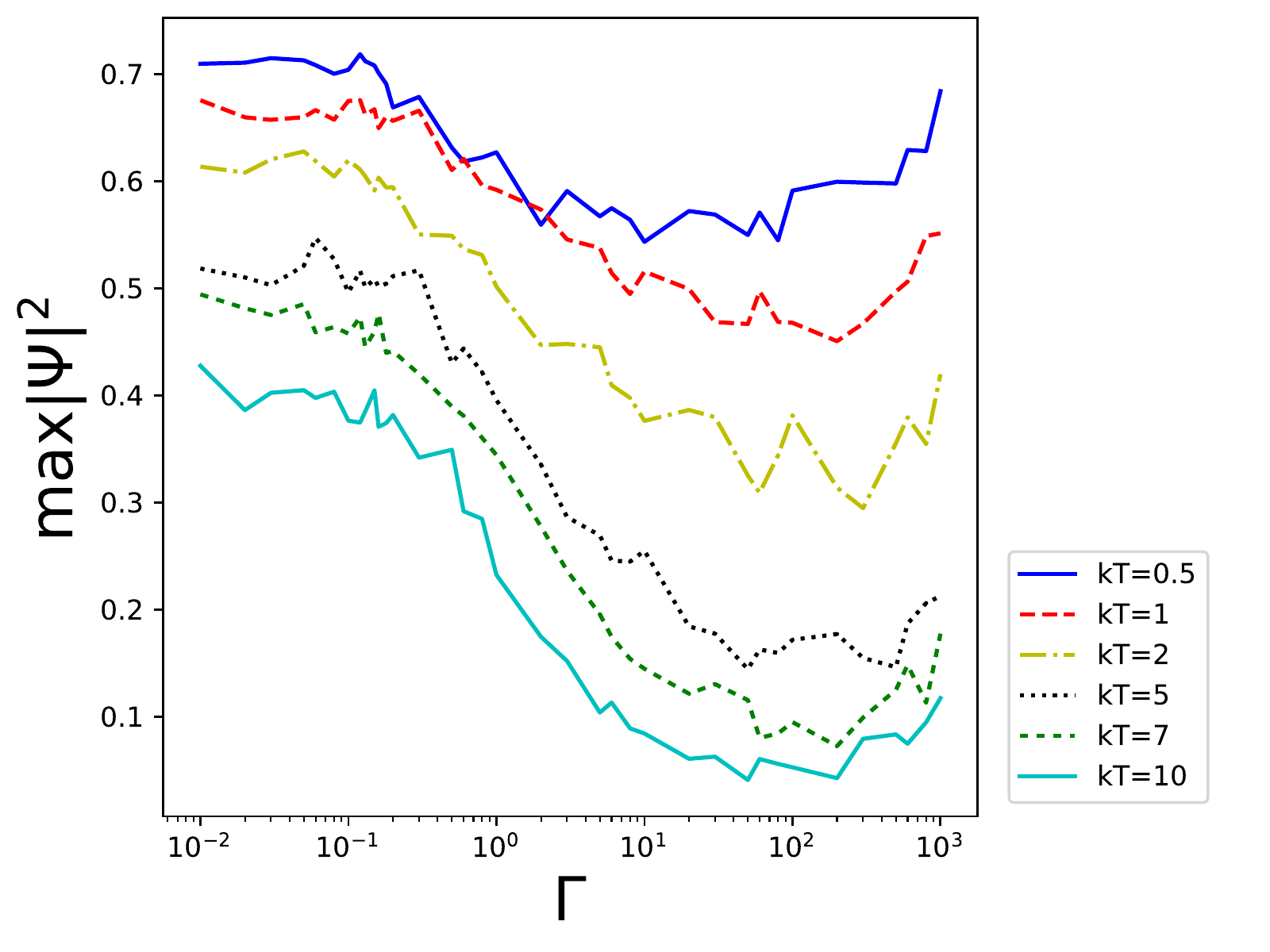}
 \includegraphics[width=80mm]{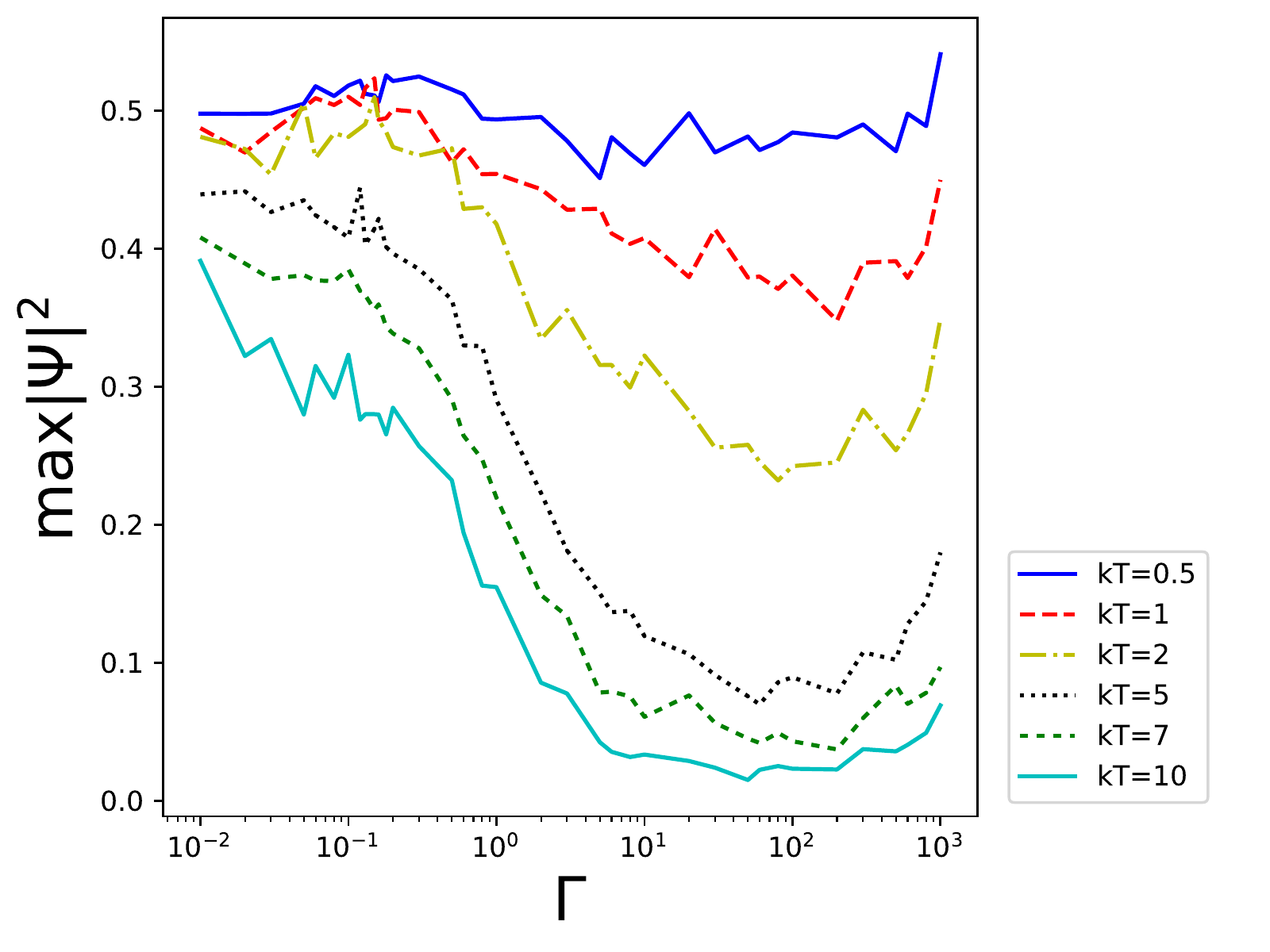}
}
\centerline{c\hspace{80mm} d }
\caption{Full homogeneous coupling. The plot of $\max |\Psi_{N+1}|^2$ for
  $\tau \le 500$ as a function of $\Gamma$ for different values of $kT$.
   a) $N=15$, b) $N=18$, c) $N=21$, d) $N=24$.
}
\label{fig:FF_gamma}
\end{figure}

In Fig. \ref{fig:FF_kt} we present the variation of $\max|\Psi_{N+1}|^2$ as a
function of temperature for different chain lengths. We see that for
short chains, the temperature has a  minimal effect while for longer
chains, its influence is more pronounced.

  It is worth to recall that although large proteins in electron
  transport chains can consist of up to a few thousand aminoacids, the
  $\alpha-$ helical parts of their globular structure consist of up to
  $N=50-70$ peptide groups. In trans-membrane proteins the $\alpha$-helices
  are even shorter, with $N=30$ or even smaller. Moreover, in the biggest
  enzymes of electron transport chain, like NADH ubiquinone oxidoreductase,
  which is the first and the biggest protein complex of the respiratory chain,
  there is a whole pathway for the electron transport prior to the
  ubiquinone reduction via several iron sulfur clusters, connected by
  relatively short $\alpha$-helices (see  \cite{DavKisl}). So we conclude,
  that
  under physiological conditions, the transfer of the electron from a donor
  to an acceptor is thermally stable.

Looking at the data in Fig. \ref{fig:FF_gamma} we see that the probability
of an electron transfer from the donor to the acceptor is
relatively constant when $\Gamma<1$ and so that it does not play a significant
role on the thermal stability of the electron transfer.
In \cite{Brizhik2014} $\Gamma$, which can only be estimated,
was taken to be $0.2$.

\section{Conclusions}
In this paper we have presented
a model describing the long-range transport of an electron from a
donor molecule to an acceptor one via the nonlinear state of a large polaron
(soliton-like state) formed in a $\alpha$-helical protein in a
`Donor -- $\alpha$-helix -- Acceptor' system. Conventionally, we model the
$\alpha$-helix as a polypeptide chain, twisted in a helix, in
which each peptide group is coupled to its nearest neighbours by a
chemical bond and to every 3\textsuperscript{rd} neighbour by a
hydrogen bond. The helix can thus be described as 3 parallel
strands coupled to each other. We have found that the static polaron on such
a helix, for the parameters that describe AMID I vibration in
$\alpha$-helical protein, is a relatively broad localized hump extended over
the polypeptide macromolecule in agreement with other studies
(see, e.g., \cite{DavErSerg, Brizhik2004}). In our model we have only taken
into account one phonon mode, while in real proteins there are many other
phonon modes, the interaction with which results in bigger value of the
effective electron-lattice coupling, and, hence, in stronger soliton
localization than obtained here. 

We have then studied the transfer of an electron from a donor molecule
to the acceptor by initially placing the electron on the donor. For
the proper parameters of the couplings, the electron was, within a very
short time interval, transferred onto the polypeptide
chain where it was self-trapped in a polaron state, and then moved
towards the other extremity of the chain where it was absorbed by the
acceptor.

We have considered three types of couplings between the donor and the
polypeptide chain as well as between the acceptor and the polypeptide
chain.  In the first case, the donor and the acceptor where coupled, 
respectively, to the first 3 and the last 3 nodes of the chain, using the
identical parameters and we called such a configuration the `fully
homogeneous' one. In the second configuration, the donor was coupled
to the first node of the chain and the acceptor to the last node of
the same strand or to the last node of the helix. We called such couplings
`single-strand' and 'end-to-end' ones, respectively. 

The fully homogeneous coupling is the one that leads to the best
donor-acceptor electron transport with an efficiency of  90\% or more
depending on the length of the chain.  The `end-to-end' coupling did
not work so well, but still led to a transfer probability of up to 60\% while
the `single-strand' one was the worst leading only to a 20\% probability
transfer. These results can be explained from the dependence of the
efficiency of the soliton generation not only on the actual parameter values
of the system, but also on the type of couplings between the helix and the
donor and the acceptor. If one uses the inverse scattering theory for integrable
system, applied to the time evolution of certain initial conditions for the
nonlinear Schr\"odinger equation that approximates Davydov solitons
\cite{Dav-Br-gen, LSB-gen}, these different couplings translate into
specific initial conditions which leads to families of solitons
with different efficiencies.

Our study has shown that an electron in the polaron (soliton-like)
state can easily propagate as a travelling wave  along the
$\alpha$-helical chain. The polaron that is generated in the helix
in the vicinity of the donor molecule has a complex
internal structure: it is not just a clean simple polaron localized on a
single strand; instead, it is distributed between the strands and propagates
along the helix with some intrinsic oscillations, rather than along a
particular strand, which reflects its collective hybrid nature. 

Unfortunately, the exact value of the exchange interaction for extra electron
in proteins is not known, but can be roughly estimated at 0.05 - 0.1 eV,
comparable with $\Jb$, and the other parameters are the same as for AMID I
vibration, which were used in our model. Other related manufactured
‘Donor - $\alpha$-helix – Acceptor’ systems, described in the introduction,
have parameter values close to the considered here, so we can conclude that
our model for the long-range electron transport describes these systems as well.

Our results explain the experimental evidence that the donor and acceptor
parameters, as well as the type of their coupling  affect the electron
transport in 'Donor -- $\alpha$-helix -- Acceptor' systems (see \cite{Jiang, Guo}).

We have also shown that when we add thermal fluctuations to the model,
the long range electron transfer in the `Donor -- $\alpha$-helix -- Acceptor'
system is stable at physiological temperatures. We have thermalized a mixed
quantum classical system in a way which  makes the quantum part  behave
‘classically’ as well. According to \cite{Cruzeiro} this results in a
broadening of the soliton wave function in the helix and in a decrease of the
binding energy of the soliton. This, in turn, results in a lower stability of
the soliton with respect to any perturbation, including thermal fluctuations,
compared to a proper analysis of the thermal stability.
Moreover,  as shown in \cite{Pershko88}, accounting for temperature fluctuations in the
equation for lattice displacements within a quantum-mechanical description
results in an effective decrease of the resonant interaction energy by the
exponential Debay-Waller factor. This then leads to a decrease of the
spatial dispersion of the electron and an increase of the electron-lattice
coupling which itself results in an increase of the binding energy of the
soliton and, as a result, a higher thermal stability compared to our model.
A proper analysis of thermal stability of electron transport would require a
more rigorous treatment and be the topic of a paper on its own. In this paper
we have decided to restrict ourselves to the simplest analysis.

\section{Acknowledgement}
One of us, LSB, acknowledges the partial support from the budget program KPKVK
6541230 and the scientific program 0117U00236 of the Department of Physics and
Astronomy of the National Academy of Sciences of Ukraine and thanks the
Department of Mathematical Sciences of the University of Durham for the
hospitality during her short-term visit. WJZ thanks the Leverhulme Trust
for his grant EM-2016-007.


\bibliographystyle{unsrt}

\end{document}